\documentclass[preprint,prd,nofootinbib,tightenlines,groupedaddress,superscriptaddress,amsmath,amssymb]{revtex4}

\usepackage{graphicx}
\usepackage{bm}
\usepackage{color}
\usepackage[hypertex]{hyperref}

\begin{document}
\baselineskip=15pt \parskip=4pt

\vspace*{3em}

\title{Probing Leptonic Interactions of a Family-Nonuniversal $\bm{Z'}$ Boson}

\author{Cheng-Wei Chiang$^{a,b,c}$, Yi-Fan Lin$^a$, and Jusak Tandean$^a$}
\affiliation{$^a$Department of Physics and Center for Mathematics and Theoretical Physics,
National Central University, Chungli 320, Taiwan
\bigskip \\
$^b$Institute of Physics, Academia Sinica, \\ Taipei 119, Taiwan
\bigskip \\
$^c$Physics Division, National Center for Theoretical Sciences, \\ Hsinchu 300, Taiwan \\
$\vphantom{\bigg|_{\bigg|}^|}$}


\begin{abstract}
We explore a $Z'$ boson with family-nonuniversal couplings to charged leptons.
The general effect of $Z$-$Z'$ mixing, of both kinetic and mass types, is included in the analysis.
Adopting a model-independent approach, we perform a comprehensive study of constraints on
the leptonic $Z'$ couplings from currently available experimental data on a number of
flavor-conserving and flavor-changing transitions.
Detailed comparisons are made to extract the most stringent bounds on the leptonic couplings.
Such information is fed into predictions of various processes that may be experimentally
probed in the near future.
\end{abstract}

\maketitle

\section{Introduction}

Recent anomalous measurements of a number of observables at the Fermilab Tevatron, such as
the forward-backward asymmetry in top-quark pair production~\cite{Abazov:2007qb},
the like-sign dimuon charge asymmetry in semileptonic $b$-hadron decays~\cite{Abazov:2010hv},
and the invariant mass distribution of jet pairs produced in association with
a~$W$~boson~\cite{Aaltonen:2011mk}, give us possible hints on physics beyond the standard model~(SM).
One of the candidates that have been proposed to explain these anomalies is a~massive
spin-one electrically neutral gauge particle, the $Z'$ boson, which may be associated with
an additional Abelian gauge symmetry, U(1)$'$, that is broken at around the TeV scale and has
a~mass of {\footnotesize$\,\sim$}150 GeV~\cite{Jung:2009jz,Deshpande:2010hy,Buckley:2011vc,Yu:2011cw}.
Moreover, the desired $Z'$ boson would need to have sufficiently sizable flavor-changing
neutral-current~(FCNC) interactions in the quark sector.

One way to induce $Z'$-mediated FCNC's is to introduce exotic fermions having U(1)$'$ charges
different from those of the SM fermions~\cite{Nardi:1992nq},
as occurs in models with the E$_6$ grand unified group.
In this case, the mixing of the right-handed ordinary and exotic quarks, all SU(2)$_L$ singlets,
gives rise to FCNC's mediated by a heavy $Z'$ or due to small $Z$-$Z'$ mixing.
Another possibility involves family-nonuniversal interactions of the~$Z'$.
In string-inspired model building, it is natural for at least one of the gauge bosons of the extra
U(1) groups to possess family-nonuniversal couplings to ordinary fermions~\cite{Chaudhuri:1994cd}.
In this scenario, the FCNC couplings appear when one transforms the SM fermions into their mass
eigenstates, without the necessity to introduce new fermion states.
Furthermore, both left- and right-handed fermions can have significant flavor-violating
interactions with the $Z'$, as well as small family-nondiagonal couplings to
the $Z$ boson caused by $Z$-$Z'$ mixing.

In fact, $Z'$ models with tree-level quark FCNC's have been studied extensively in low-energy
flavor physics phenomena, such as neutral meson ($K$, $D$, or $B$) mixing, $B$-meson decays
involving the \,$b\to s$\, transition in particular, and single top
production~\cite{Langacker:2000ju,Barger:2004qc,Chiang:2006we}.
In principle, one can consider the possibility of FCNC's in the lepton sector as
well~\cite{Langacker:2000ju,arXiv:1107.5238}.
In this work, we focus on family-nonuniversal interactions of the $Z'$ with the charged leptons
and explore constraints on its relevant couplings from various experiments involving only leptons
in the initial and final states.
Such processes suffer less from QCD corrections and hadronic uncertainties than
the above-mentioned hadronic systems.
We assume that the $Z'$ boson arises from an extra U(1) gauge symmetry,
but otherwise adopt a model-independent approach.
We take into account the effect of $Z$-$Z'$ mixing, of both kinetic and mass types,
which modifies theoretical predictions of the electroweak $\rho$ parameter and various
$Z$-pole observables.
Due to the family nonuniversality, such a~$Z'$ boson would feature flavor-changing leptonic
couplings, as would also the $Z$ boson through the mixing.
We therefore examine a~number of flavor-conserving and flavor-changing processes to evaluate
constraints on the leptonic $Z'$ couplings.

This paper is organized as follows. We present the interactions of the $Z'$ boson with
the charged leptons in Section~\ref{sec:interactions}.
The $\rho$ parameter from global electroweak fits is used to determine the allowed mixing angle
between the~$Z$ and~$Z'$.
In Section~\ref{sec:flavorconserving}, we study constraints on the flavor-conserving couplings
of the~$Z'$.
The pertinent observables include those in leptonic $Z$ decays from the $Z$-pole data and
the cross sections of \,$e^+e^-$\, collisions into lepton-antilepton pairs measured at LEP\,II.
We separate the analysis of the flavor-changing couplings into two parts.
The constraints from transitions generated by tree-level diagrams are treated
in Section~\ref{sec:flavorchanging}.
We place upper bounds on the couplings from the rates of flavor-violating \,$Z\to\bar ll'$\, decays,
\,$\mu\to3e$,\, several flavor-violating $\tau$ decays into 3 leptons, muonium-antimuonium
conversion, as well as the cross sections of flavor-changing annihilations~\,$e^+e^-\to\bar l l'$.\,
The constraints from processes induced by loop diagrams are given
in Section~\ref{sec:loopflavorchanging}.
The considered processes or observables are the flavor-changing radiative lepton decays
\,$l\to l'\gamma$,\, the anomalous magnetic moments of leptons, and their electric dipole moments.
We will make use of the existing experimental information on all these transitions,
including new measurements from the BaBar, Belle, and MEG
Collaborations~\cite{Aubert:2009tk,Hayasaka:2010np,Adam:2011ch}.
Based on the allowed coupling ranges, we make predictions for various flavor-conserving and
-violating processes in Section~\ref{sec:predictions}.
These predictions can serve to help guide experimentalists in future searches for $Z'$ signals.
Our findings are summarized in Section~\ref{sec:summary}.

\section{Interactions \label{sec:interactions}}

The mass Lagrangian for the interaction eigenstates $\hat Z$ and $\hat Z'$
of the massive neutral gauge bosons after electroweak symmetry breaking,
which leaves the photon massless, can be expressed~as
\begin{eqnarray}
{\cal L}_{\rm m}^{} \,\,=\,\,
\frac{1}{2}\bigl(\hat Z^\lambda~~~\hat Z^{\prime\lambda}\bigr)
\left(\begin{array}{ccc} {M}_Z^2 && \Delta \vspace{1ex} \\
\Delta  && {M}_{Z'}^2 \end{array}\right)
\left(\begin{array}{c} \hat Z_\lambda^{} \vspace{1ex} \\
\hat Z_\lambda' \end{array}\right) ~,
\end{eqnarray}
where $M_{Z,Z'}^{}$ denote the masses of the gauge bosons and $\Delta$ represents the mixing
between them.
As discussed in Appendix~\ref{lag}, which has some more details on the notation we adopt,
$\Delta$~contains both possible kinetic- and mass-mixing contributions, and in the presence of
kinetic mixing the parameter $M_{Z'}^{}$ is not identical to the original mass of the U(1)$'$
gauge boson [see~Eq.\,(\ref{mz'})].

The squared-mass matrix in ${\cal L}_{\rm m}^{}$ can be diagonalized using~\cite{Langacker:2008yv}
\begin{eqnarray}
\left(\begin{array}{c} \hat Z \vspace{1ex} \\ \hat Z' \end{array}\right) =
\left(\begin{array}{ccc} \cos\xi^{} && -\sin\xi^{} \vspace{1ex} \\
\sin\xi^{} && \cos\xi^{} \end{array}\!\right) \!
\left(\begin{array}{c} Z \vspace{1ex} \\ Z' \end{array}\right) , \hspace{5ex}
\tan(2\xi) \,=\, \frac{2\Delta}{{M}_Z^2-{M}_{Z'}^2} ~,
\end{eqnarray}
with its eigenvalues being
\begin{eqnarray} \label{mzmz'}
m_{Z,Z'}^2 \,=\, \mbox{$\frac{1}{2}$}\bigl({M}_Z^2+{M}_{Z'}^2\bigr) \mp
\mbox{$\frac{1}{2}$}\sqrt{\bigl({M}_Z^2-{M}_{Z'}^2\bigr)^2+4\Delta^2} ~.
\end{eqnarray}
One can then derive
\begin{eqnarray}   \label{tan2x}
\bigl(m_{Z'}^2-{M}_Z^2\bigr)\tan^2\xi \,\,=\,\, {M}_Z^2-m_Z^2 ~.
\end{eqnarray}

The Lagrangian describing the interactions of $\hat Z$ and $\hat Z'$ with the charged leptons is
\begin{eqnarray}   \label{L_int}
{\cal L}_{\rm int}^{} \,\,=\,\,
-g_Z^{} J_Z^\lambda\, \hat Z_\lambda^{} \,-\, g_{Z'}^{} J_{Z'}^\lambda\, \hat Z_\lambda' ~,
\end{eqnarray}
and the currents are given by
\begin{eqnarray}
g_Z^{}\,J_Z^\lambda \,\,=\,\,
\overline{\hat{\ell}\,}\gamma^\lambda\bigl(g_L^{}P_L^{}+g_R^{}P_R^{}\bigr)\hat\ell ~,
\hspace{5ex}
g_{Z'}^{}\,J_{Z'}^\lambda \,\,=\,\,
\overline{\hat\ell\,}\gamma^\lambda\bigl(g_L'P_L^{}+g_R'P_R^{}\bigr)\hat\ell ~,
\end{eqnarray}
where \,$\hat\ell=(\hat e~~\hat\mu~~\hat\tau)^{\rm T}$\, contains the interaction eigenstates
of the leptons, \,$P_{L,R}^{}=\frac{1}{2}(1\mp\gamma_5^{})$,\, and the coupling constants
$g_{L,R}^{}$ are family universal, whereas the $\hat Z'$ couplings are not assumed to be
family universal according to
\begin{eqnarray}
g_L' \,\,=\,\, {\rm diag}\bigl(L_e',L_\mu',L_\tau'\bigr) ~, \hspace{5ex}
g_R' \,\,=\,\, {\rm diag}\bigl(R_e',R_\mu',R_\tau'\bigr) ~,
\end{eqnarray}
with the parameters \,$L_{e,\mu,\tau}'$\, and \,$R_{e,\mu,\tau}'$\, being generally different
from one another.
The Hermiticity of ${\cal L}_{\rm int}$ requires these coupling constants to be real.
The interaction eigenstates in $\hat\ell$ are related to the mass eigenstates in
\,$\ell=(e~~\mu~~\tau)^{\rm T}$\, by\footnote{Throughout the paper
we make a distinction between $\ell$ and $l$, with the former referring to the triplet of charged
leptons and the latter to individual charged leptons in general.}
\begin{eqnarray}
\hat\ell_L^{} \,\,=\,\, P_L^{}\,\hat\ell \,\,=\,\, V_L^{}\,\ell_L^{} ~, \hspace{5ex}
\hat\ell_R^{} \,\,=\,\, P_R^{}\,\hat\ell \,\,=\,\, V_R^{}\,\ell_R^{} ~,
\end{eqnarray}
where $V_{L,R}^{}$ are unitary matrices which diagonalize the lepton mass matrix $\hat M_\ell$
in the Yukawa Lagrangian,
\,${\rm diag}\bigl(m_e^{},m_\mu^{},m_\tau^{}\bigr)=V_L^\dagger\hat M_\ell^{}V_R^{}$.\,

In terms of the mass eigenstates, $Z$, $Z'$, and $\ell$, we can then write
\begin{eqnarray}   \label{Lint}
{\cal L}_{\rm int}^{} &=&
-\bar\ell\gamma^\lambda\bigl[ \bigl(g_L^{}\,\cos\xi+B_L^{}\,\sin\xi\bigr)P_L^{} +
\bigl(g_R^{}\,\cos\xi+B_R^{}\,\sin\xi\bigr)P_R^{}\bigr]\ell\,Z_\lambda^{}
\nonumber \\ && \!-\;
\bar\ell\gamma^\lambda\bigl[ \bigl(-g_L^{}\,\sin\xi+B_L^{}\,\cos\xi\bigr)P_L^{} +
\bigl(-g_R^{}\,\sin\xi+B_R^{}\,\cos\xi\bigr)P_R^{}\bigr]\ell\,Z_\lambda'
\nonumber \\ &=&
-\bar\ell_i^{}\gamma^\lambda\Bigl(\beta_L^{\ell_i\ell_j}P_L^{} +
\beta_R^{\ell_i\ell_j}P_R^{}\Bigr)\ell_j^{}\,Z_\lambda^{} \,-\,
\bar\ell_i^{}\gamma^\lambda\Bigl(b_L^{\ell_i\ell_j}P_L^{} +
b_R^{\ell_i\ell_j}P_R^{}\Bigr)\ell_j^{}\,Z_\lambda' ~,
\end{eqnarray}
where \,$B_L^{}=V_L^\dagger\,g_L'V_L^{}$\, and
\,$B_R^{}=V_R^\dagger\,g_R'V_R^{}$\, are generally nondiagonal 3$\times$3 matrices,
summation over \,$i,j=1,2,3$\, is implied, \,$\ell_{1,2,3}=e,\mu,\tau$,\, and
\begin{eqnarray} \label{beta,b}
\beta_{\sf C}^{\ell_i\ell_j} \,\,=\,\, \bigl(\beta_{\sf C}^{\ell_j\ell_i}\bigr)^* \,\,=\,\,
\delta_{ij}^{}\,c_\xi^{}\,g_{\sf C}^{} +
s_\xi^{}\,\bigl(B_{\sf C}^{}\bigr)_{ij} ~, \hspace{5ex}
b_{\sf C}^{\ell_i\ell_j} \,\,=\,\, \bigl(b_{\sf C}^{\ell_j\ell_i}\bigr)^* \,\,=\,\,
-\delta_{ij}^{}\,s_\xi^{}\,g_{\sf C}^{} + c_\xi^{}\,\bigl(B_{\sf C}^{}\bigr)_{ij}
\end{eqnarray}
for \,${\sf C}=L$ or $R$,\, with \,$c_\xi^{}=\cos\xi$\, and \,$s_\xi^{}=\sin\xi$.\,
One can see from Eq.\,(\ref{Lint}) that the presence of nonzero off-diagonal elements of
$B_{L,R}^{}$, due to the nonuniversality of the diagonal elements of $g_{L,R}'$ and
to the charged-lepton mixing, gives rise to flavor-changing couplings of
the $Z'$ to the leptons at tree level.
Furthermore, $Z$-$Z'$ mixing introduces not only family nonuniversality, but also
flavor violation into the tree-level interactions of~the~$Z$.

\begin{figure}[hbt]
\includegraphics[width=100mm]{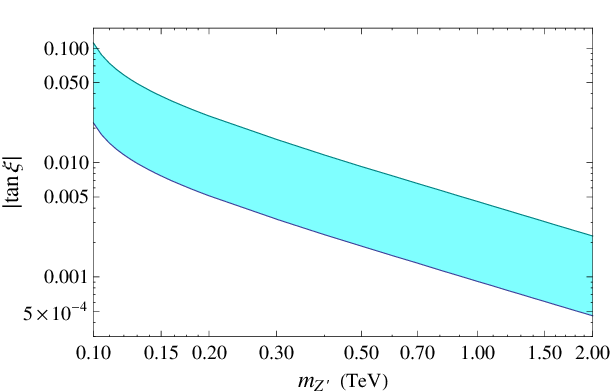}\vspace*{-1ex}
\caption{Values of \,$|$$\tan\xi$$|$\, for \,$100{\rm\,GeV}\le m_{Z'}^{}\le2$\,TeV\,
corresponding to the $\rho_0^{}$ range from the electroweak global fit.\label{tanx}}
\end{figure}

Now, it follows from Eq.\,(\ref{beta,b}) that
\begin{eqnarray}   \label{betab}
\beta_{\sf C}^{\ell_i\ell_j} \,\,=\,\,
\delta_{ij}^{}\,\frac{g_{\sf C}^{}}{c_\xi^{}} \,+\, t_\xi^{}\,b_{\sf C}^{\ell_i\ell_j} ~,
\end{eqnarray}
where \,$t_\xi^{}=\tan\xi$.\,
Therefore the couplings of $Z$ and $Z'$ to $\bar\ell_i^{}\ell_j^{}$ are directly related once
the mixing angle $\xi$ is specified.
Employing the electroweak data, one can fix  $\xi$ if the $Z'$ mass is given.
We achieve this by means of the $\rho_0^{}$ parameter, which in the Particle Data
Group~(PDG) convention~\cite{erler-pdg} encodes the effects of new physics if it
deviates from the SM expectation~\,$\rho_0^{\rm SM}=m_W^2/\bigl(c_{\rm w}^2M_Z^2\bigr)=1$,
where $c_{\rm w}^{}$ is the cosine of the Weinberg angle~$\theta_W$.
Since $Z$-$Z'$ mixing alters the $Z$ mass, as indicated in Eq.\,(\ref{tan2x}), and hence
causes $\rho_0^{}$ to shift from unity, we have
\begin{eqnarray} \label{rho0}
\rho_0^{} \,\,=\,\, \frac{m_W^2}{c_{\rm w}^2\,m_Z^2} \,\,=\,\,
\frac{m_W^2}{c_{\rm w}^2\,{M}_Z^2}
\Biggl[ 1 \,-\, \frac{m_{Z'}^2-{M}_Z^2}{{M}_Z^2}\,\tan^2\xi \Biggr]^{-1}
\,\,\simeq\,\, 1 \,+\, \frac{m_{Z'}^2-m_Z^2}{m_Z^2}\,\xi^2 ~.
\end{eqnarray}
The value \,$\rho_0^{}=1.0008^{+0.0017}_{-0.0007}$\,~\cite{erler-pdg} resulting from
the PDG global electroweak fit then translates for \,$m_{Z'}^{}=150$\,GeV\, into
\begin{eqnarray}
0.008 \,\,\le\,\, |\xi| \,\,\le\,\, 0.038 ~.
\end{eqnarray}
More generally, Fig.\,\ref{tanx} shows the corresponding limits of $|\!\tan\xi|$ for
\,$100{\rm\,GeV}\le m_{Z'}^{}\le2$\,TeV,\, which is the range of interest in this paper.
It is then straightforward to realize that for this mass range
\begin{eqnarray}
\frac{m_{Z'}^2}{m_Z^2}\,\tan^2\xi \,\,\ll\,\, 1 ~.
\end{eqnarray}
The plot also illustrates that \,$|$$\tan\xi|\propto1/m_{Z'}^{}$\, as $m_{Z'}^{}$ becomes large,
which reflects the relation
\begin{eqnarray} \label{tanxi}
|\tan\xi| \,\,\simeq\,\, \frac{m_Z^{}}{m_{Z'}^{}}\sqrt{\frac{\rho_0^{}-1}{\rho_0^{}}}
\end{eqnarray}
valid for \,$m_{Z'}^2\gg m_Z^2$\, and derived from Eq.\,(\ref{rho0}).

\section{Flavor-conserving couplings of $\bm{Z'}$\label{sec:flavorconserving}}

With $\xi$ known, one can evaluate $b_{L,R}^{ll}$ from the $Z$-pole data.
The amplitude for the $Z$ decay into a charged-lepton pair $l^+l^-$ is
\begin{eqnarray} \label{MZ2ll}
{\cal M}_{Z\to l^+l^-}^{} \,\,=\,\, \bar l\gamma_\lambda^{}\bigl(\beta_L^{ll}P_L^{} +
\beta_R^{ll}P_R^{}\bigr)l\,\varepsilon_Z^\lambda
\end{eqnarray}
in the parametrization of Eq.\,(\ref{Lint}).
This leads to the forward-backward asymmetry at the $Z$ pole and decay rate
\begin{eqnarray}   \label{GZ2ll}
A_{\rm FB}^{(0,l)} \,\,=\,\, \frac{3}{4}\,A_e^{}A_l^{} ~, \hspace{5ex}
\Gamma_{Z\to l^+l^-}^{} \,\,=\,\, \frac{\sqrt{m_Z^2-4m_l^2}}{16\pi\,m_Z^2}\;
\overline{\bigl|{\cal M}_{Z\to l^+l^-}^{}\bigr|^2} ~,
\end{eqnarray}
where
\begin{eqnarray}   \label{Al}
A_l^{} \,\,=\,\,
\frac{\bigl(\beta_L^{ll}\bigr)^2-\bigl(\beta_R^{ll}\bigr)^2}
{\bigl(\beta_L^{ll}\bigr)^2+\bigl(\beta_R^{ll}\bigr)^2} ~, \hspace{5ex}
\overline{\bigl|{\cal M}_{Z\to l^+l^-}^{}\bigr|^2} \,\,=\,\,
\frac{2}{3}\Bigl[\bigl(\beta_L^{ll}\bigr)^2+\bigl(\beta_R^{ll}\bigr)^2\Bigr]
\bigl(m_Z^2-m_l^2\bigr)
\,+\, 4\,m_l^2\,\beta_L^{ll}\beta_R^{ll} ~. ~~~
\end{eqnarray}

These formulas along with Eq.\,(\ref{betab}) allow us to extract $b_{L,R}^{ll}$ for each value
of $\xi$ from the $A_l^{}$ and $\Gamma_{Z\to l^+l^-}^{}$ measurements~\cite{pdg},
\begin{eqnarray} \label{zpole} &
A_e^{\rm exp} \,\,=\,\, 0.1515\pm0.0019 ~, ~~~~ A_\mu^{\rm exp} \,\,=\,\, 0.142\pm0.015 ~, ~~~~
A_\tau^{\rm exp} \,\,=\,\, 0.143\pm0.004 ~, & \nonumber \\ &
\Gamma_{Z\to e^+e^-}^{\rm exp} \,\,=\,\, 83.91\pm0.12 {\rm\;MeV} ~, ~~~~
\Gamma_{Z\to\mu^+\mu^-}^{\rm exp} \,\,=\,\, 83.99\pm0.18 {\rm\;MeV} ~, & \\ &
\Gamma_{Z\to\tau^+\tau^-}^{\rm exp} \,\,=\,\, 84.08\pm0.22 {\rm\;MeV} ~, & ~~~ \nonumber
\end{eqnarray}
after $g_{L,R}^{}$ are fixed from their SM predictions~\cite{erler-pdg}
\begin{eqnarray}
A_e^{\rm SM} &=& A_\mu^{\rm SM} \,\,=\,\, A_\tau^{\rm SM} \,\,=\,\, 0.1475\pm0.0010 ~,
\nonumber \\ \Gamma_{Z\to e^+e^-}^{\rm SM} \,\,=\,\, \Gamma_{Z\to\mu^+\mu^-}^{\rm SM} &=&
84.00\pm0.06 {\rm\;MeV} ~, \hspace{5ex}
\Gamma_{Z\to\tau^+\tau^-}^{\rm SM} \,\,=\,\, 83.82\pm0.06 {\rm\;MeV} ~.
\end{eqnarray}
We can reproduce all these SM numbers within their errors using Eqs.~(\ref{GZ2ll}) and
(\ref{Al}) with $\beta_L^{ll}$ and $\beta_R^{ll}$ replaced, respectively,
by the effective couplings
\begin{eqnarray}   \label{geff}
g_L^{\rm eff} \,\,=\,\, -0.1996 ~, \hspace{5ex} g_R^{\rm eff} \,\,=\,\,  0.1721 ~.
\end{eqnarray}
For comparison, their tree-level values are
\,$g_L^{}=g\bigl(s_{\rm w}^2-1/2\bigr)/c_{\rm w}^{}\simeq-0.2002$\, and
\,$g_R^{}=g s_{\rm w}^2/c_{\rm w}^{}\simeq0.1722$\, if \,$s_{\rm w}^2=0.23116$\,~\cite{pdg}.
We will ignore the uncertainties in $g_{L,R}^{\rm eff}$ compared to the greater relative
uncertainties in the data.

\begin{figure}[b] \vspace{2ex}
\includegraphics[width=57mm]{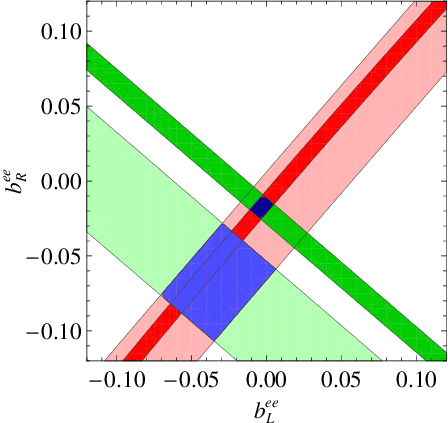}
\hfill
\includegraphics[width=56mm]{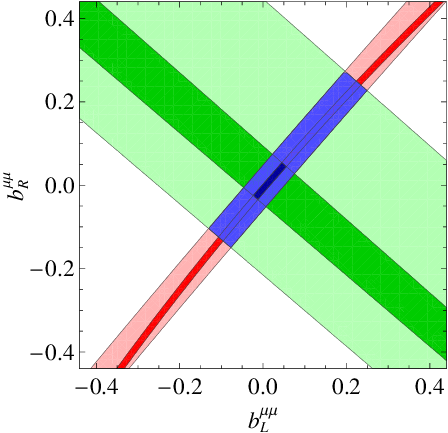}
\hfill
\includegraphics[width=56mm]{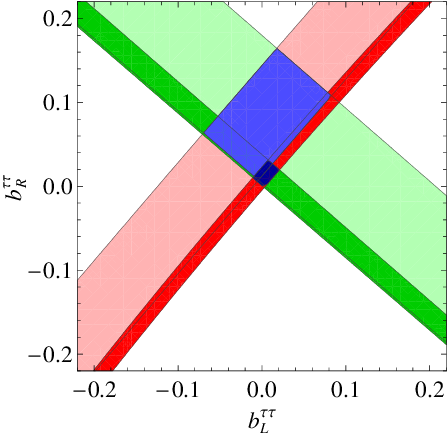}
\vspace*{-1ex}
\caption{Values of $b_{L,R}^{ll}$ for \,$m_{Z'}^{}=150$\,GeV\, and mixing angle
\,$\xi=0.008$~(lighter colors), $0.038$~(darker colors),\, as described in the text,
subject to constraints from $A_l^{}$ and $\Gamma_{Z\to l^+l^-}$ data.\label{bLbR}}
\end{figure}

Applying \,$\beta_{L,R}^{ll}=g_{L,R}^{}/c_\xi^{}+t_\xi^{}\,b_{L,R}^{ll}$\, from
Eq.\,(\ref{betab}) in the $A_l^{}$ and $\Gamma_{Z\to l^+l^-}^{}$ formulas above,
with \,$g_{L,R}^{}=g_{L,R}^{\rm eff}$\, from Eq.\,(\ref{geff}) and
a specific value of~$\xi$, one can then obtain the allowed ranges of $b_{L,R}^{ll}$
from the measured values in Eq.\,(\ref{zpole}) within their one-sigma ranges.
Thus in the \,$m_{Z'}^{}=150$\,GeV\, case, for which \,$0.008\le|\xi|\le0.038$,\,
the results for \,$\xi>0$\, are
\begin{eqnarray} \label{bLbR150} &
-0.071 \,\,\le\,\, b_L^{ee} \,\,\le\,\, 0.006 ~, \hspace{5ex}
-0.11 \,\,\le\,\, b_R^{ee} \,\,\le\,\, -0.009 ~, & \nonumber \\ &
-0.13 \,\,\le\,\, b_L^{\mu\mu} \,\,\le\,\, 0.25 ~, \hspace{5ex}
-0.15 \,\,\le\,\, b_R^{\mu\mu} \,\,\le\,\, 0.27 ~, & \nonumber \\ &
-0.070 \,\,\le\,\, b_L^{\tau\tau} \,\,\le\,\, 0.083 ~, \hspace{5ex}
-0.002 \,\,\le\,\, b_R^{\tau\tau} \,\,\le\,\, 0.16 ~. & ~~~~~
\end{eqnarray}
Flipping the sign of $\xi$ would also flip the signs of these $b_{L,R}^{ll}$ numbers,
and the same statement applies to the rest of our analysis.
The plots in Fig.\,\ref{bLbR} illustrate the allowed $b_{L,R}^{ll}$ regions for
the lower and upper limits of the $\xi$ range in this case, \,$\xi=0.008$ (lighter colors) and
\,$\xi=0.038$ (darker colors).\,
The green regions satisfy the $A_l^{}$ constraints, red the $\Gamma_{Z\to l^+l^-}$
constraints, and blue both of them.
The upper and lower limits of the $b_{L,R}^{ll}$ ranges in Eq.\,(\ref{bLbR150}) are visible on
the plots.

Since $Z'$-mediated diagrams can also affect the collision \,$e^+e^-\to l^+l^-$,\, it is important
to consider the relevant data to see if they offer additional restraints on the $Z'$ couplings.
Here we will employ LEP-II measurements at various center-of-mass energies above
the $Z$ pole, from 130 to~207~GeV\,~\cite{Alcaraz:2006mx}.
The amplitude for this process if \,$l\neq e$\, is
\begin{eqnarray} \label{Mee2ll}
{\cal M}_{e^+ e^-\to\bar l l}^{} &=&
-\frac{e_p^2\,\bar l\gamma^\nu l\, \bar e\gamma_\nu^{}e}{s} \,+\,
\frac{\bar l\gamma^\nu\bigl(\beta_L^{ll}P_L^{}+\beta_R^{ll}P_R^{}\bigr)l\,
\bar e\gamma_\nu^{}\bigl(\beta_L^{ee}P_L^{}+\beta_R^{ee}P_R^{}\bigr)e}{m_Z^2-s}
\nonumber \\ && \!+\;
\frac{\bar l\gamma^\nu\bigl(b_L^{ll}P_L^{}+b_R^{ll}P_R^{}\bigr)l\,
\bar e\gamma_\nu^{}\bigl(b_L^{ee}P_L^{}+b_R^{ee}P_R^{}\bigr)e}{m_{Z'}^2-s} ~,
\end{eqnarray}
where \,$e_p^{}>0$\, is the proton's electric charge, \,$s=(p_{e^+}+p_{e^-})^2$,\, and
we have assumed that $s$ is not near $m_{Z,Z'}^2$.
There are also contributions to this amplitude from $t$-channel diagrams with
flavor-changing couplings~$\beta_{L,R}^{e l}$ and~$b_{L,R}^{e l}$, but we will neglect their
effects in order to explore the largest impact of the $Z'$ flavor-conserving couplings under
the assumption that there is no unnatural cancellation between the two sets of contributions.
Moreover, as we demonstrate below, the magnitudes of the latter couplings have looser upper-limits
than their flavor-changing counterparts by at least a~few times.
Complete expressions for the cross-section $\sigma(e^+e^-\to l^+l^-)$ and forward-backward
asymmetry $A_{\rm FB}$, including finite-width effects, are collected in~Appendix~\ref{csafb}.
Numerically, we adopt the couplings in Eq.\,(\ref{geff}) and the effective value~\,$\alpha=1/132$,\,
which in the absence of the $Z'$ lead to $\sigma$ and $A_{\rm FB}$ numbers differing by
no more than 2~percent from the corresponding SM predictions quoted in the LEP-II
report~\cite{Alcaraz:2006mx}.
Since we consider \,$m_{Z'}^{}=150$~GeV and larger masses from 0.5 to~2~TeV,\,
in determining the $g_{L,R}^{ll}$ bounds we take the LEP-II data belonging to
\,$\sqrt s=136,161,205,207$~GeV\, for definiteness.

We find that incorporating the LEP-II information brings about significant modifications to
some of the results in~Eq.\,(\ref{bLbR150}).
The allowed values of the couplings for \,$m_{Z'}^{}=150$\,GeV\, now become
\begin{eqnarray} \label{bLbR150new} &
-0.071 \,\,\le\,\, b_L^{ee} \,\,\le\,\, 0.006 ~, \hspace{5ex}
 -0.10 \,\,\le\,\, b_R^{ee} \,\,\le\,\, -0.009 ~, & \nonumber \\ &
-0.033 \,\,\le\,\, b_L^{\mu\mu} \,\,\le\,\, 0.080 ~, \hspace{5ex}
-0.029 \,\,\le\,\, b_R^{\mu\mu} \,\,\le\,\, 0.095 ~, & \nonumber \\ &
-0.070 \,\,\le\,\, b_L^{\tau\tau} \,\,\le\,\, 0.024 ~, \hspace{10ex}
     0 \,\,\le\,\, b_R^{\tau\tau} \,\,\le\,\, 0.083 ~. & ~~~~~
\end{eqnarray}

We have also explored the situations for higher masses up to \,$m_{Z'}^{}=2$\,TeV.\,
The inclusion of the LEP-II data again provide important extra restrictions on
the couplings.\footnote{This also occurs in the case of family-universal $Z'$ studied in
Ref.\,\cite{delAguila:2010mx}. The bounds on the leptonic couplings found therein are
roughly comparable to ours.}
For the representative values \,$m_{Z'}^{}=0.5$~-~2~TeV,\, the allowed ranges associated with
each flavor turn out to be roughly proportional to the $m_{Z'}^{}$ values, namely
\begin{eqnarray} \label{fcc} & \displaystyle
-5.1 \,\,\lesssim\,\, \frac{b_L^{ee}}{m_{Z'}^{}} \,\,\lesssim\,\, -1.2 ~, \hspace{5ex}
-5.4 \,\,\lesssim\,\, \frac{b_R^{ee}}{m_{Z'}^{}} \,\,\lesssim\,\, -1.1 ~, & \nonumber
\\ & \displaystyle
-4.3 \,\,\lesssim\,\, \frac{b_L^{\mu\mu}}{m_{Z'}^{}} \,\,\lesssim\,\, 3.4 ~, \hspace{6.5ex}
-4.3 \,\,\lesssim\,\, \frac{b_R^{\mu\mu}}{m_{Z'}^{}} \,\,\lesssim\,\, 2.1 ~, & \nonumber
\\ & \displaystyle
-6.1 \,\,\lesssim\,\, \frac{b_L^{\tau\tau}}{m_{Z'}^{}} \,\,\lesssim\,\, -2.0 ~, \hspace{7ex}
1.9 \,\,\lesssim\,\, \frac{b_R^{\tau\tau}}{m_{Z'}^{}} \,\,\lesssim\,\, 5.9 ~, & ~~
\end{eqnarray}
where the numbers are in units of \,$10^{-4}$~GeV$^{-1}$.\,
In obtaining all the $b_{L,R}^{ll}$ ranges above, we let the couplings be present
at the same time.
It is worth noting that the proportionality of these ranges to the $Z'$ mass for
\,$m_{Z'}^{}\gg m_Z^{}$\, is a reflection of the \,$|$$\tan\xi|\propto1/m_{Z'}^{}$\,
behavior in~Eq.\,(\ref{tanxi}) which starts to manifest itself when $m_{Z'}^{}$ exceeds
200\,GeV or so, as can be seen in Fig.\,\ref{tanx}.
We also note that for \,$m_{Z'}^{}\,{\scriptstyle\gtrsim}\,2$\,TeV\, the limits in Eq.\,(\ref{fcc})
accommodate couplings which may exceed order one in magnitude and hence the perturbativity limit.
Nevertheless, as the errors in $\rho_0^{}$ decrease with increasingly better precision in future
data, the bounds on $b_{L,R}^{ll}$ will likely become stronger.

Before proceeding to the flavor-changing sector, a few comments regarding the case of no
$Z$-$Z'$ mixing, \,$\xi=0$,\, are in order.
If one goes beyond the one-sigma range of the $\rho_0^{}$ parameter from the global
electroweak fit, so that the lower bound of $\rho_0^{}$ reaches zero,
then the lower bound of $|\xi|$ will also reach zero.
In that limit \,$\beta_{L,R}^{ll}\to g_{L,R}^{}$,\, and therefore the $Z$-pole data on
$A_l^{}$ and $\Gamma_{Z\to l^+l^-}$ no longer offer restrictions on $b_{L,R}^{ll}$
through the tree-level relations in Eqs.\,(\ref{GZ2ll}) and~(\ref{Al}).
At the one-loop level, however, $Z'$-mediated radiative corrections contribute to
the $Zl^+l^-$ vertex, and so these observables can still constrain
the couplings~\cite{Carone:1994aa}.
With the formulas given in Ref.\,\cite{Carone:1994aa} for the $Z'$-loop contribution,
we estimate that the upper limits on the coupling-to-mass ratios, $b/m_{Z'}^{}$, are of order
1 to 2 per mill GeV$^{-1}$ for our $m_{Z'}^{}$ range of interest and thus higher than their
counterparts in the presence of mixing.
Without the mixing, $Z'$-mediated diagrams can still affect \,$e^+e^-\to l^+l^-$\, at
tree level, as Eq.\,(\ref{Mee2ll}) indicates.
The expressions for the cross section and forward-backward asymmetry in Appendix~\ref{csafb}
suggest, however, that the LEP-II data would not impose additional restrictions in this case.

\section{Constraints from tree-level flavor-changing processes \label{sec:flavorchanging}}

\subsection{$\bm{Z\to e^\pm\mu^\mp}$, \,$\bm{Z\to e^\pm\tau^\mp}$,\, and
\,$\bm{Z\to\mu^\pm\tau^\mp}$}

As ${\cal L}_{\rm int}^{}$ in Eq.\,(\ref{Lint}) shows, the $Z$ can have
tree-level flavor-violating interactions with leptons in the presence of $Z$-$Z'$ mixing.
Accordingly, the amplitude of the decay \,$Z\to l\bar l'$\, for \,$l'\neq l$\, is
\begin{eqnarray}
{\cal M}_{Z\to l\bar l'}^{} \,\,=\,\, \bar l\gamma_\lambda^{}\bigl(\beta_L^{ll'}P_L^{} +
\beta_R^{ll'}P_R^{}\bigr)l'\,\varepsilon_Z^\lambda ~,
\end{eqnarray}
where \,$\beta_{L,R}^{ll'}=t_\xi^{}\,b_{L,R}^{ll'}$\, from Eq.\,(\ref{betab}).
The rate of this transition is then
\begin{eqnarray}
\Gamma_{Z\to l\bar l'}^{} \,\,=\,\, \frac{|\bm{p}_l^{}|\,t_\xi^2}{8\pi m_Z^2} \Biggl\{ \!
\Bigl(\bigl|b_L^{ll'\!}\bigr|^2+\bigl|b_R^{ll'\!}\bigr|^2\Bigr)
\Biggl[\frac{2m_Z^2-m_l^2-m_{l'}^2}{3}-\frac{(m_l^2-m_{l'}^2)^2}{3m_Z^2}\Biggr] +
4 m_l^{}m_{l'}^{}\, {\rm Re}\bigl(b_L^{ll'\!*}b_R^{ll'}\bigr) \biggr\} \,, ~~~
\end{eqnarray}
where $\bm{p}_l^{}$ is the three-momentum of $l$  in the $Z$ rest-frame.
These decays, like all other lepton-flavor-violating ones, have not yet been observed.
But there is some experimental information available on the branching ratios:
\,${\cal B}(Z\to e^\pm\mu^\mp)<1.7\times10^{-6}$,\,
\,${\cal B}(Z\to e^\pm\tau^\mp)<9.8\times10^{-6}$,\, and
\,${\cal B}(Z\to\mu^\pm\tau^\mp)<1.2\times10^{-5}$\,~\cite{pdg},
each of the numbers being the sum of contributions from the listed final states.
Assuming that only one of $\beta_{L,R}^{ll'}$ is nonzero at a time, we can then obtain
constraints on $b_{L,R}^{ll'}$ after specifying $\xi$ associated with a given $Z'$ mass.
Thus for \,$m_{Z'}^{}=150$\,GeV,\, in which case \,$0.008\le|\xi|\le0.038$,\,
\begin{eqnarray}
\bigl|b_{L,R}^{e\mu}\bigr| \,\,\le\,\, 0.17 ~, \hspace{5ex}
\bigl|b_{L,R}^{e\tau}\bigr| \,\,\le\,\, 0.41 ~, \hspace{5ex}
\bigl|b_{L,R}^{\mu\tau}\bigr| \,\,\le\,\, 0.44 ~.
\end{eqnarray}
For the higher masses, \,$m_{Z'}^{}=0.5$~-~2~TeV,\, we find that the upper bounds are again
approximately proportional to their masses,
\begin{eqnarray}
\frac{\bigl|b_{L,R}^{e\mu}\bigr|}{m_{Z'}^{}} \,\,\lesssim\,\, 1.4 ~, \hspace{5ex}
\frac{\bigl|b_{L,R}^{e\tau}\bigr|}{m_{Z'}^{}} \,\,\lesssim\,\, 3.5 ~, \hspace{5ex}
\frac{\bigl|b_{L,R}^{\mu\tau}\bigr|}{m_{Z'}^{}} \,\,\lesssim\,\, 3.8
\end{eqnarray}
in units of \,$10^{-3}$~GeV$^{-1}$.\,
Stricter constraints can come from some of the other processes we study in the following.

\subsection{$\bm{\mu\to3e}$, \,$\bm{\tau\to3e}$,\, and \,$\bm{\tau\to 3\mu}$}

The decay \,$\mu^-\to e^-e^+e^-$\, receives contributions from diagrams involving
the $Z$ and~$Z'$.
From ${\cal L}_{\rm int}^{}$ in Eq.~(\ref{Lint}), we derive the amplitude for the tree-level
contributions to be
\begin{eqnarray}
{\cal M}_{\mu\to3e}^{} &=&
\biggl( \frac{\beta_L^{ee}\beta_L^{e\mu}}{m_Z^2} +
\frac{b_L^{ee}b_L^{e\mu}}{m_{Z'}^2} \biggr)
\bar e\gamma^\nu P_L^{}e\,\bar e'\gamma_\nu^{} P_L^{}\mu
+ \biggl( \frac{\beta_R^{ee}\beta_R^{e\mu}}{m_Z^2} +
\frac{b_R^{ee}b_R^{e\mu}}{m_{Z'}^2} \biggr)
\bar e\gamma^\nu P_R^{}e\,\bar e'\gamma_\nu^{} P_R^{}\mu
\nonumber \\ && \! +\;
\biggl( \frac{\beta_L^{ee}\beta_R^{e\mu}}{m_Z^2} +
\frac{b_L^{ee}b_R^{e\mu}}{m_{Z'}^2} \biggr)
\bar e\gamma^\nu P_L^{}e\,\bar e'\gamma_\nu^{} P_R^{}\mu
+ \biggl( \frac{\beta_R^{ee}\beta_L^{e\mu}}{m_Z^2} +
\frac{b_R^{ee}b_L^{e\mu}}{m_{Z'}^2} \biggr)
\bar e\gamma^\nu P_R^{}e\,\bar e'\gamma_\nu^{} P_L^{}\mu
\nonumber \\ && \! -\; \bigl(\bar e\leftrightarrow\bar e'\bigr) ~.
\end{eqnarray}
Here we use $\bar e$ and $\bar e'$ to distinguish the two electrons in the final state.
The minus sign in the above equation comes from Fermi statistics.
Using \,$\beta_{L,R}^{e\mu}=t_\xi^{}b_{L,R}^{e\mu}$\, and ignoring the electron mass,
we can write the resulting branching ratio as
\begin{eqnarray}
{\cal B}(\mu\to3e) &=&
\frac{\tau_\mu^{}\,m_\mu^5}{1536\,\pi^3}\Biggl\{
\Biggl[2\biggl(\frac{t_\xi^{}\beta_L^{ee}}{m_Z^2}+\frac{b_L^{ee}}{m_{Z'}^2}\biggr)^{\!\!2}\!+
\biggl(\frac{t_\xi^{}\beta_R^{ee}}{m_Z^2}+\frac{b_R^{ee}}{m_{Z'}^2}\biggr)^{\!\!2} \Biggr]
\bigl|b_L^{e\mu}\bigr|^2
\nonumber \\ && \hspace*{9ex} +\;
\Biggl[ \biggl(\frac{t_\xi^{}\beta_L^{ee}}{m_Z^2}+\frac{b_L^{ee}}{m_{Z'}^2}\biggr)^{\!\!2}\!+
2\biggl(\frac{t_\xi^{}\beta_R^{ee}}{m_Z^2}+\frac{b_R^{ee}}{m_{Z'}^2}\biggr)^{\!\!2} \Biggr]
\bigl|b_R^{e\mu}\bigr|^2 \Biggr\} ~,
\end{eqnarray}
where $\tau_\mu^{}$ is the $\mu$ lifetime and
\,$\beta_{L,R}^{ee}=g_{L,R}^{}/c_\xi^{} + t_\xi^{}\, b_{L,R}^{ee}$.\,

To evaluate the upper limits on $|b_{L,R}^{e\mu}|{}^2$ from the data on~\,$\mu\to3e$,\,
one can try to look for nonzero minima of the coefficients of $|b_{L,R}^{e\mu}|^2$ in
the ${\cal B}(\mu\to3e)$ formula.
After scanning the values of $\xi$ and $b_{L,R}^{ee}$ satisfying the experimental
requirements discussed in the previous section, we find for \,$m_{Z'}^{}=150$\,GeV\, that
the minimum of the coefficient of \,$|b_L^{e\mu}|{}^2$\, is \,$4.7\times10^{-4}$\, at
\,$\bigl(b_L^{ee},b_R^{ee},\xi\bigr)\simeq\pm(0.0042,-0.021,0.025)$,\, whereas that of
\,$|b_R^{e\mu}|{}^2$\, is \,$3.1\times10^{-4}$\, at
\,$\bigl(b_L^{ee},b_R^{ee},\xi\bigr)\simeq\pm(0.0045,-0.018,0.031)$.\,
From the measured bound
\,${\cal B}(\mu^-\to e^-e^+e^-)_{\rm exp}^{}<1.0\times10^{-12}$\,~\cite{pdg},
assuming as before that only one of $\beta_{L,R}^{e\mu}$ is nonvanishing at a time,
we then extract in the \,$m_{Z'}^{}=150$\,GeV\, case
\begin{eqnarray} \label{em1}
\bigl|b_L^{e\mu}\bigr| \,\,\le\,\, 4.6\times10^{-5} ~, \hspace{5ex}
\bigl|b_R^{e\mu}\bigr| \,\,\le\,\, 5.7\times10^{-5} ~.
\end{eqnarray}
For \,$m_{Z'}^{}=0.5$~-~2~TeV,\, taking similar steps we obtain the limits to be roughly
proportional to $m_{Z'}^{}$ according to
\begin{eqnarray} \label{em2}
\frac{\bigl|b_L^{e\mu}\bigr|}{m_{Z'}^{}} \,\,\lesssim\,\, 1.4\times10^{-7}{\rm~GeV}^{-1} ~,
\hspace{5ex}
\frac{\bigl|b_R^{e\mu}\bigr|}{m_{Z'}^{}} \,\,\lesssim\,\, 1.8\times10^{-7}{\rm~GeV}^{-1} ~.
\end{eqnarray}

In the analogous case of \,$\tau^-\to e^-e^+e^-$,\, the expression for the branching ratio
can be simply derived from that for~\,${\cal B}(\mu\to3e)$\, by replacing each $\mu$ in
the indices with~$\tau$.
The same can be said about the coefficients of $\bigl|b_{L,R}^{e\tau}\bigr|{}^2$
in the ${\cal B}(\tau\to3e)$ formula.
It follows that the measured bound
\,${\cal B}(\tau^-\to e^-e^+e^-)_{\rm exp}^{}<2.7\times10^{-8}$\,~\cite{pdg}
yields for \,$m_{Z'}^{}=150$\,GeV\,
\begin{eqnarray} \label{et1}
\bigl|b_L^{e\tau}\bigr| \,\,\le\,\, 0.018 ~, \hspace{5ex}
\bigl|b_R^{e\tau}\bigr| \,\,\le\,\, 0.022 ~,
\end{eqnarray}
whereas for \,$m_{Z'}^{}=0.5$~-~2~TeV\,
\begin{eqnarray} \label{et2}
\frac{\bigl|b_L^{e\tau}\bigr|}{m_{Z'}^{}} \,\,\lesssim\,\, 5.3\times10^{-5}{\rm~GeV}^{-1} ~,
\hspace{5ex}
\frac{\bigl|b_R^{e\tau}\bigr|}{m_{Z'}^{}} \,\,\lesssim\,\, 6.9\times10^{-5}{\rm~GeV}^{-1} ~.
\end{eqnarray}

As for \,$\tau^-\to\mu^-\mu^+\mu^-$,\, upon scanning the allowed values of
$b_{L,R}^{\mu\mu}$ and $\xi$ we find that the coefficients of
$\bigl|b_{L,R}^{\mu\tau}\bigr|{}^2$ in the ${\cal B}(\tau\to3\mu)$ formula have minima
which are vanishingly small.
Consequently, this mode cannot provide useful restraints on~$\bigl|b_{L,R}^{\mu\tau}\bigr|$
separately.

\subsection{$\bm{\tau\to\mu\bar e e}$\, and \,$\bm{\tau\to e\bar\mu\mu}$}

Another transition that can happen in our $Z'$ scenario is~\,$\tau^-\to\mu^-e^+e^-$.\,
The tree-level contribution to its amplitude is
\begin{eqnarray}
{\cal M}_{\tau\to\mu\bar e e}^{} &=&
\biggl( \frac{\beta_L^{ee}\beta_L^{\mu\tau}}{m_Z^2} +
\frac{b_L^{ee}b_L^{\mu\tau}}{m_{Z'}^2} \biggr)
\bar e\gamma^\nu P_L^{}e\,\bar\mu\gamma_\nu^{} P_L^{}\tau
+ \biggl( \frac{\beta_L^{ee}\beta_R^{\mu\tau}}{m_Z^2} +
\frac{b_L^{ee}b_R^{\mu\tau}}{m_{Z'}^2} \biggr)
\bar e\gamma^\nu P_L^{}e\,\bar\mu\gamma_\nu^{} P_R^{}\tau
\nonumber \\ && \! -\;
\biggl( \frac{\beta_L^{\mu e}\beta_L^{e\tau}}{m_Z^2} +
\frac{b_L^{\mu e}b_L^{e\tau}}{m_{Z'}^2} \biggr)
\bar\mu\gamma^\nu P_L^{}e\,\bar e\gamma_\nu^{} P_L^{}\tau
- \biggl( \frac{\beta_L^{\mu e}\beta_R^{e\tau}}{m_Z^2} +
\frac{b_L^{\mu e}b_R^{e\tau}}{m_{Z'}^2} \biggr)
\bar\mu\gamma^\nu P_L^{}e\,\bar e\gamma_\nu^{} P_R^{}\tau
\nonumber \\ && \! +\; (L\leftrightarrow R) ~.
\end{eqnarray}
It leads to the branching ratio
\begin{eqnarray} \label{Btmee}
{\cal B}(\tau\to\mu\bar e e) &=&
\frac{\tau_\tau^{}\,m_\tau^5}{1536\pi^3}\Biggl[
\biggl|\biggl(\frac{t_\xi^{}\beta_L^{ee}}{m_Z^2}+\frac{b_L^{ee}}{m_{Z'}^2}\biggr)b_L^{\mu\tau} +
\frac{b_L^{\mu e}b_L^{e\tau}}{m_{Z'}^2}\biggr|^2 +
\biggl|\biggl(\frac{t_\xi^{}\beta_R^{ee}}{m_Z^2}+\frac{b_R^{ee}}{m_{Z'}^2}\biggr)b_R^{\mu\tau} +
\frac{b_R^{\mu e}b_R^{e\tau}}{m_{Z'}^2}\biggr|^2
\nonumber \\ && \hspace*{8ex} +\,
\biggl(\frac{t_\xi^{}\beta_L^{ee}}{m_Z^2}+\frac{b_L^{ee}}{m_{Z'}^2}\biggr)^{\!\!2}
|b_R^{\mu\tau}|^2 +
\biggl(\frac{t_\xi^{}\beta_R^{ee}}{m_Z^2}+\frac{b_R^{ee}}{m_{Z'}^2}\biggr)^{\!\!2}
|b_L^{\mu\tau}|^2 +
\frac{|b_L^{\mu e}b_R^{e\tau}|^2 +
|b_R^{\mu e}b_L^{e\tau}|^2}{m_{Z'}^2} \Biggr] , \nonumber \\
\end{eqnarray}
where  final lepton masses have been neglected and terms containing
\,$\bigl|\beta_{\sf C}^{\mu e}\beta_{\sf C'}^{e\tau}\bigr|=
t_\xi^2\bigl|b_{\sf C}^{\mu e}b_{\sf C'}^{e\tau}\bigr|$\,
for \,${\sf C,C'}=L,R$\, have been dropped because \,$t_\xi^2\ll m_Z^2/m_{Z'}^2$.\,
To determine the upper bounds on~\,$|b_{L,R}^{\mu\tau}|^2$,\, one can again then try to seek
nonvanishing minima of their coefficients in~Eq.\,(\ref{Btmee}) which are
the same, under the assumption that $b_{\sf C}^{\mu e,e\tau}$ are absent.
Thus for \,$m_{Z'}^{}=150$\,GeV\, we place the minimum to be \,$4.9\times10^{-5}$\, at
\,$\bigl(b_L^{ee},b_R^{ee},\xi\bigr)\simeq\pm(0.0043,-0.019,0.028)$.\,
From the experimental information
\,${\cal B}(\tau^-\to\mu^-e^+e^-)_{\rm exp}^{}<1.8\times10^{-8}$\,~\cite{pdg},
we subsequently extract for \,$m_{Z'}^{}=150$\,GeV\,
\begin{eqnarray} \label{mt1}
\bigl|b_{L,R}^{\mu\tau}\bigr| \,\,\le\,\, 0.019 ~.
\end{eqnarray}
Similarly, for \,$m_{Z'}^{}=0.5$~-~2~TeV\,  we arrive at
\begin{eqnarray} \label{mt2}
\frac{\bigl|b_{L,R}^{\mu\tau}\bigr|}{m_{Z'}^{}}\,\,\lesssim\,\,6\times10^{-5}{\rm~GeV}^{-1} ~.
\end{eqnarray}
Assuming \,$b_{L,R}^{\mu\tau}=0$\, instead, we get
\begin{eqnarray}
\frac{\bigl|b_{\sf C}^{\mu e}b_{\sf C'}^{e\tau}\bigr|}{m_{Z'}^2}
\,\,\le\,\, 1.0\times10^{-8} {\rm~GeV}^{-2} ~.
\end{eqnarray}
The constraints in the last equation are weaker by \,{\footnotesize$\sim$}3\, orders of
magnitude than those put together from~Eqs.~(\ref{em1})-(\ref{et2}).

For \,$\tau^-\to e^-\mu^+\mu^-$,\, the expression for the branching ratio follows from that
for~\,${\cal B}(\tau\to\mu\bar e e)$\, with $e$ and $\mu$ being interchanged in the indices.
In this case the coefficients of $\bigl|b_{L,R}^{e\tau}\bigr|{}^2$ in
\,${\cal B}(\tau\to e\bar\mu\mu)$\, have vanishingly small minima.
Hence useful upper-bounds on these couplings are not available from
\,${\cal B}(\tau^-\to e^-\mu^+\mu^-)_{\rm exp}^{}<2.7\times10^{-8}$\,~\cite{pdg}.
On the other hand, assuming \,$b_{L,R}^{e\tau}=0$\, we can extract
\begin{eqnarray}
\frac{\bigl|b_{\sf C}^{e\mu}b_{\sf C'}^{\mu\tau}\bigr|}{m_{Z'}^2}
\,\,\le\,\, 1.3\times10^{-8} {\rm~GeV}^{-2} ~,
\end{eqnarray}
which are also very weak compared to what can be deduced from~Eqs.~(\ref{em2}) and~(\ref{mt2}).

\subsection{$\bm{\tau\to e e\bar\mu}$\, and \,$\bm{\tau\to\bar e\mu\mu}$}

Like the preceding ones, the \,$\tau^-\to e^-e^-\mu^+$\, decay receives tree-level contributions
proceeding from~Eq.\,(\ref{Lint}), but involves two flavor-changing vertices exclusively.
The amplitude is given by
\begin{eqnarray}
{\cal M}_{\tau\to e e\bar\mu}^{} &=&
\biggl(\frac{\beta_L^{e\mu}\beta_L^{e\tau}}{m_Z^2}+\frac{b_L^{e\mu}b_L^{e\tau}}{m_{Z'}^2}\biggr)
\bar e\gamma^\nu P_L^{}\mu\,\bar e'\gamma_\nu^{} P_L^{}\tau +
\biggl(\frac{\beta_R^{e\mu}\beta_R^{e\tau}}{m_Z^2}+\frac{b_R^{e\mu}b_R^{e\tau}}{m_{Z'}^2}\biggr)
\bar e\gamma^\nu P_R^{}\mu\,\bar e'\gamma_\nu^{} P_R^{}\tau
\nonumber \\ && \! +\;
\biggl(\frac{\beta_L^{e\mu}\beta_R^{e\tau}}{m_Z^2}+\frac{b_L^{e\mu}b_R^{e\tau}}{m_{Z'}^2}\biggr)
\bar e\gamma^\nu P_L^{}\mu\,\bar e'\gamma_\nu^{} P_R^{}\tau +
\biggl(\frac{\beta_R^{e\mu}\beta_L^{e\tau}}{m_Z^2}+\frac{b_R^{e\mu}b_L^{e\tau}}{m_{Z'}^2}\biggr)
\bar e\gamma^\nu P_R^{}\mu\,\bar e'\gamma_\nu^{} P_L^{}\tau
\nonumber \\ && \! -\; \bigl(\bar e\leftrightarrow\bar e'\bigr) ~.
\end{eqnarray}
Neglecting the terms involving \,$\beta_{\sf C}^{e\mu}\beta_{\sf C'}^{e\tau}$\, as before,
we consequently have
\begin{eqnarray}
{\cal B}(\tau\to e e\bar\mu) &=&
\frac{\tau_\tau^{}\,m_\tau^5}{1536\,\pi^3}\Biggl(
\frac{2|b_L^{e\mu}|^2+|b_R^{e\mu}|^2}{m_{Z'}^4}\bigl|b_L^{e\tau}\bigr|^2 +
\frac{|b_L^{e\mu}|^2+2|b_R^{e\mu}|^2}{m_{Z'}^4}\bigl|b_R^{e\mu}\bigr|^2 \Biggr) ~.
\end{eqnarray}
The measurement
\,${\cal B}(\tau^-\to\mu^+e^-e^-)_{\rm exp}^{}<1.5\times10^{-8}$\,~\cite{pdg} then implies
\begin{eqnarray}
\frac{\bigl|{b}_{L,R}^{e\mu}{b}_{L,R}^{e\tau}\bigr|}{m_{Z'}^2} \,\,\le\,\, 6.8\times10^{-9}
~\mbox{GeV}^{-2} ~,
\hspace{5ex}
\frac{\bigl|{b}_{L,R}^{e\mu}{b}_{R,L}^{e\tau}\bigr|}{m_{Z'}^2} \,\,\le\,\, 9.6\times10^{-9}
~\mbox{GeV}^{-2} ~.
\end{eqnarray}
For \,$\tau^-\to e^+\mu^-\mu^-$,\, following analogous steps we obtain from
\,${\cal B}(\tau^-\to e^+\mu^-\mu^-)_{\rm exp}^{}<1.7\times10^{-8}$\,~\cite{pdg} that
\begin{eqnarray}
\frac{\bigl|{b}_{L,R}^{\mu e}{b}_{L,R}^{\mu\tau}\bigr|}{m_{Z'}^2} \,\,\le\,\, 7.2\times10^{-9}
~\mbox{GeV}^{-2} ~,
\hspace{5ex}
\frac{\bigl|{b}_{L,R}^{\mu e}{b}_{R,L}^{\mu\tau}\bigr|}{m_{Z'}^2} \,\,\le\,\, 1.0\times10^{-8}
~\mbox{GeV}^{-2} ~.
\end{eqnarray}
All these results are again less strict than the corresponding constraints inferred from
Eqs.~(\ref{em2}), (\ref{et2}), and~(\ref{mt2}) by roughly 3 orders of magnitude.

\subsection{Muonium-antimuonium conversion \,$\bm{\mu^+e^-\to\mu^-e^+}$}

The experimental information on \,$\mu^+e^-\to\mu^-e^+$\, is available in terms of
the effective parameter~$G_C^{}$ which is defined by~\cite{pdg,Willmann:1998gd}
\begin{eqnarray}
{\cal L}_{\rm eff}^{} \,\,=\,\, \sqrt{8}\,G_C^{}\,\bar\mu\gamma^\nu P_{\sf C'\,}^{}e\,
\bar\mu\gamma_\nu^{} P_{{\sf C}'\,}^{}e \;+\; {\rm H.c.} ~,
\end{eqnarray}
with \,${\sf C}'=L$ or $R$,\,
and has been measured to be \,$|G_C^{}|<0.0030\,G_{\rm F}^{}$~\cite{pdg}, where
$G_{\rm F}^{}$ is the Fermi coupling constant.
Attributing this to the $Z'$ implies that
\begin{eqnarray} \label{gc}
\frac{\bigl|b_{L,R}^{\mu e}\bigr|}{m_{Z'}^{}} \,\,=\,\, 2\sqrt{\sqrt{2}\,|G_C^{}|}
\,\,\le\,\, 4.4\times10^{-4}{\rm~GeV}^{-1} ~,
\end{eqnarray}
far less restrictive than~Eq.\,(\ref{em2}).

\subsection{Flavor violating \,$\bm{e^+e^-\to\bar l l'}$}

At \,$e^+e^-$\, colliders, new physics could trigger the production of flavor-violating events
with $e\mu$, $e\tau$, and $\mu\tau$ in the final states.
In our $Z'$ scenario, the tree-level amplitude of \,$e^+e^-\to l^+l^{\prime-}$\, for
\,$l'\neq l$\, is
\begin{eqnarray}
{\cal M}_{\bar e e\to\bar l l'}^{} &=&
\frac{\bar l'\gamma^\nu\bigl(\beta_L^{l'l}P_L^{}+\beta_R^{l'l}P_R^{}\bigr)l\,
\bar e\gamma_\nu^{}\bigl(\beta_L^{ee}P_L^{}+\beta_R^{ee}P_R^{}\bigr)e}{m_Z^2-s}
\nonumber \\ && -\;
\frac{\bar e\gamma^\nu\bigl(\beta_L^{e l}P_L^{}+\beta_R^{e l}P_R^{}\bigr)l\,
\bar l'\gamma_\nu^{}\bigl(\beta_L^{l'e}P_L^{}+\beta_R^{l'e}P_R^{}\bigr)e}{m_Z^2-t}
\nonumber \\ && +\; \bigl(Z\to Z',\,\beta\to b\bigr) ~,
\end{eqnarray}
where \,$s=\bigl(p_{e^+}^{}+p_{e^-}^{}\bigr){}^2$\, is assumed not to be close to $m_{Z,Z'}^2$
and \,$t=\bigl(p_{e^+}^{}-p_{l^+}^{}\bigr){}^2$.\,
The first experimental limits on the cross sections
\,$\sigma(ll')\equiv\sigma(\bar e e\to\bar l l')+\sigma(\bar e e\to l\bar l')$\,
were acquired by the OPAL Collaboration~\cite{Abbiendi:2001cs}
at LEP-II energies,~${\cal O}(200$\,GeV).\,
More recent bounds on the cross sections at much lower energies, around \,11 and
1\,GeV,\, were reported by the BaBar~\cite{Aubert:2006uy} and
SND~\cite{Achasov:2009en} Collaborations, respectively.
Since the theoretical cross sections tend to grow significantly as the energy increases
from 1 to 200~GeV, the OPAL data~\cite{Abbiendi:2001cs}
\,$\bar\sigma(e\mu)_{\rm exp}^{}<22$\,fb,\, \,$\bar\sigma(e\tau)_{\rm exp}^{}<78$\,fb,\, and
\,$\bar\sigma(\mu\tau)_{\rm exp}^{}<64$\,fb\, for the average cross sections over
\,$200{\rm\,GeV}\le\sqrt s\le209$\,GeV\, impose potentially stronger restraints than the others.
The cross sections at these energies being more sensitive to the effects of
\,$m_{Z'}^{}=150$\,GeV\, than to those of \,$m_{Z'}^{}\ge0.5$\,TeV,\,
we discuss only the case of the former, in which for \,$l=\mu$ or $\tau$\,
\begin{eqnarray}
\bar\sigma(el) &\simeq& \bigl[
38.4\,\bigl(\beta_L^{ee}\,t_\xi\bigr){}^2 + 175\, \bigl(\beta_R^{ee}\,t_\xi\bigr){}^2
- 20.9\,\beta_L^{ee}\,b_L^{ee}\,t_\xi + 230\,\beta_R^{ee}\,b_R^{ee}\,t_\xi
\nonumber \\ && ~+\,
9.68\, \bigl(b_L^{ee}\bigr){}^2 + 103\, \bigl(b_R^{ee}\bigr){}^2 \bigr] \bigl|b_L^{l e}\bigr|{}^2
\times10^4{\rm~fb}
\nonumber \\ && +\; (L\leftrightarrow R) ~,
\\ \vphantom{|^{\big|}}
\bar\sigma(\mu\tau) &\simeq& \bigl\{
\bigl[ 193\, \bigl(\beta_L^{ee}\,t_\xi\bigr){}^2 + 669\, \beta_L^{ee}\,b_L^{ee}\, t_\xi +
581\, \bigl(b_L^{ee}\bigr){}^2 \,+\, (L\leftrightarrow R) \bigr] \bigl|b_L^{\mu\tau}\bigr|{}^2
\nonumber \\ && ~-\,
\bigl( 413\, \beta_L^{ee}\, t_\xi + 717\, b_L^{ee} \bigr)\,
{\rm Re}\bigl(b_L^{e\mu}b_L^{\mu\tau}b_L^{\tau e}\bigr)
\,+\,
\bigl(233\,\bigl|b_L^{e\mu}\bigr|{}^2+448\,\bigl|b_R^{e\mu}\bigr|{}^2\bigr)
\bigl|b_L^{\tau e}\bigr|{}^2 \bigr\} \times10^3{\rm~fb}
\nonumber \\ && +\; (L\leftrightarrow R) ~. ~~~~
\end{eqnarray}
Minimizing the coefficients of $\bigl|b_{L,R}^{l e}\bigr|{}^2$ in $\bar\sigma(el)$
and comparing the latter to their data then yields
\begin{eqnarray} & \displaystyle
\bigl|b_L^{\mu e}\bigr| \,\,<\,\, 0.76 ~, \hspace{5ex}
\bigl|b_R^{\mu e}\bigr| \,\,<\,\, 0.52 ~, \hspace{5ex}
\bigl|b_L^{\tau e}\bigr| \,\,<\,\, 1.4 ~, \hspace{5ex}
\bigl|b_R^{\tau e}\bigr| \,\,<\,\, 1.0 ~. &
\end{eqnarray}
In an analogous way, $\bar\sigma(\mu\tau)$ in the absence of $b_{L,R}^{e\mu}$
gives \,$\bigl|b_{L,R}^{\mu\tau}\bigr|\le 1.2$,\,
whereas assuming \,$b_{L,R}^{\mu\tau}=0$\, instead leads to
\begin{eqnarray}
\bigl|b_{L,R}^{e\mu}b_{L,R}^{\tau e}\bigr| \,\,\le\,\, 0.017 ~, \hspace{5ex}
\bigl|b_{L,R}^{e\mu}b_{R,L}^{\tau e}\bigr| \,\,\le\,\, 0.012 ~.
\end{eqnarray}
These are all weaker than their counterparts from Eqs.~(\ref{em1}), (\ref{et1}),
and~(\ref{mt1}) by 50 times or more.

\section{Constraints from loop-generated processes \label{sec:loopflavorchanging}}

\subsection{$\bm{\mu\to e\gamma}$, \,$\bm{\tau\to e\gamma}$,\, and \,$\bm{\tau\to\mu\gamma}$}

The flavor-violating radiative decay \,$l\to l'\gamma$\, occurs at the loop level, and its
amplitude takes the general gauge-invariant form
\begin{eqnarray} \label{Mll'g}
{\cal M}_{l\to l'\gamma}^{} \,\,=\,\, i\varepsilon_\mu^*k_\nu^{}\,
\bar l'\bigl(\Sigma_L^{l'l}P_L^{}+\Sigma_R^{l'l}P_R^{}\bigr)\sigma^{\mu\nu}l ~,
\end{eqnarray}
where $k$ is the momentum of the outgoing photon, the parameters $\Sigma_{L,R}^{l'l}$ depend
on the loop contents, and \,$\sigma^{\nu\omega}=\frac{i}{2}[\gamma^\nu,\gamma^\omega]$.\,
This leads to the branching ratio
\begin{eqnarray} \label{l2lg}
{\cal B}(l\to l'\gamma) \,\,=\,\,
\frac{\tau_l^{}\,\bigl(m_l^2-m_{l'}^2\bigr)^3}{16\pi\,m_l^3}
\Bigl(\bigl|\Sigma_L^{l'l}\bigr|^2 + \bigl|\Sigma_R^{l'l}\bigr|^2\Bigr) ~,
\end{eqnarray}
where $\tau_l^{}$ is the $l$ lifetime.

\begin{figure}[b]
\includegraphics[width=100mm]{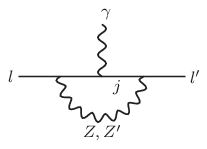}
\caption{Diagram for $Z$ and $Z'$ contributions to flavor-violating radiative
decay~\,$l\to l'\gamma$.\label{loop}}
\end{figure}

This decay receives $Z$- and $Z'$-induced contributions via the diagram displayed
in~Fig.\,\ref{loop}, with internal lepton~$j$.
Since the masses $m_{l,l',j}^{}$ of the external and internal leptons are small relative
to~$m_{Z,Z'}^{}$, it is a good approximation to retain only the lowest order terms
in expanding the loop functions in terms of $m_{l,l',j}^{}/m_{Z,Z'}^{}$.
In that limit, we can employ the results of Ref.\,\cite{He:2009rz} to derive
for negatively charged leptons
\begin{eqnarray} \label{sigmaLR} &&
\Sigma_L^{l'l} \,=\,\, \frac{e_p^{}}{24\pi^2\,m_Z^2}\sum_j\bigl(3\beta_R^{l'j}\beta_L^{jl}\,m_j^{}
- \beta_L^{l'j}\beta_L^{jl}\,m_{l'}^{} - \beta_R^{l'j}\beta_R^{jl}\,m_l^{} \bigr)
\;+\; \bigl(Z\to Z',\,\beta\to b\bigr) ~, \nonumber \\ &&
\Sigma_R^{l'l} \,=\,\, \frac{e_p^{}}{24\pi^2\,m_Z^2}\sum_j\bigl(3\beta_L^{l'j}\beta_R^{jl}\,m_j^{}
- \beta_R^{l'j}\beta_R^{jl}\,m_{l'}^{} - \beta_L^{l'j}\beta_L^{jl}\,m_l^{} \bigr)
\;+\; \bigl(Z\to Z',\,\beta\to b\bigr) ~,
\end{eqnarray}
where the sum is over \,$j=e,\mu,\tau$\, and \,$\beta_{L,R}^{jl}=\beta_{L,R}^{lj*}$.\,
Since  \,$m_\tau^{}\simeq17\,m_\mu^{}\gg m_e^{}$,\, we consider only the most enhanced terms in
$\Sigma_{L,R}^{l'l}$, the ones proportional to $m_\tau^{}$.
Accordingly
\begin{eqnarray} \label{Sem}  & \displaystyle
\Sigma_L^{e\mu} \,\,=\,\, \frac{e_p^{}\,m_\tau^{}\, b_R^{e\tau}b_L^{\tau\mu}}{8\pi^2\,m_{Z'}^2} ~,
\hspace{5ex}
\Sigma_R^{e\mu} \,\,=\,\, \frac{e_p^{}\,m_\tau^{}\,b_L^{e\tau}b_R^{\tau\mu}}{8\pi^2\,m_{Z'}^2} ~,
& \\ \nonumber \\ & \displaystyle
\Sigma_L^{e\tau} \,\,=\,\, \frac{e_p^{}\,m_\tau^{}}{24\pi^2} \Biggl[
\frac{\bigl(3\beta_L^{\tau\tau}-\beta_R^{\tau\tau}-\beta_R^{ee}\bigr)\beta_R^{e\tau}}{m_Z^2}
+ \frac{\bigl(3b_L^{\tau\tau}-b_R^{\tau\tau}-b_R^{ee}\bigr)b_R^{e\tau}
- b_R^{e\mu}b_R^{\mu\tau}}{m_{Z'}^2} \Biggr] ~,
& \nonumber \\ & \displaystyle
\Sigma_L^{\mu\tau} \,\,=\,\, \frac{e_p^{}\,m_\tau^{}}{24\pi^2} \Biggl[
\frac{\bigl(3\beta_L^{\tau\tau}-\beta_R^{\tau\tau}-\beta_R^{\mu\mu}\bigr)\beta_R^{\mu\tau}}{m_Z^2}
+ \frac{\bigl(3b_L^{\tau\tau}-b_R^{\tau\tau}-b_R^{\mu\mu}\bigr)b_R^{\mu\tau}
- b_R^{\mu e}b_R^{e\tau}}{m_{Z'}^2} \Biggr] ~, & \label{Slt}
\end{eqnarray}
and $\Sigma_R^{e\tau,\mu\tau}$ follow from $\Sigma_L^{e\tau,\mu\tau}$ with $L$ and $R$
interchanged, where we have also neglected terms with $\beta_{\sf C}^{e\mu}\beta_{\sf C}^{\mu\tau}$
$\bigl(\beta_{\sf C}^{\mu e}\beta_{\sf C}^{e\tau}\bigr)$ in
$\Sigma_{\sf C}^{e\tau}$~$\bigl(\Sigma_{\sf C}^{\mu\tau}\bigr)$.

The newest information from recent searches for these modes is
\,${\cal B}(\mu\to e\gamma)_{\rm exp}^{}<2.4\times10^{-12}$\, reported by
the MEG Collaboration~\cite{Adam:2011ch}.
With the aid of Eqs.\,(\ref{l2lg}) and~(\ref{Sem}), it translates into
\begin{eqnarray} \label{mtte}
\frac{\bigl|b_{L,R}^{\mu\tau}{b}_{R,L}^{\tau e}\bigr|}{m_{Z'}^2} \,\,\le\,\, 2.6\times10^{-11} ~.
\end{eqnarray}
These numbers are 2 orders of magnitude smaller than the corresponding ones combined from
Eqs.\,(\ref{et2}) and~(\ref{mt2}) and therefore complement them.

The present bounds for the other 2 decays are not as strong,
\,${\cal B}(\tau\to e\gamma)_{\rm exp}^{}<3.3\times10^{-8}$\, and
\,${\cal B}(\tau\to\mu\gamma)_{\rm exp}^{}<4.4\times10^{-8}$\, from BaBar~\cite{Aubert:2009tk,pdg}.
Setting \,$b_{L,R}^{e\mu}=0$\, first, one can try to evaluate from these data the biggest
$\bigl|b_{L,R}^{e\tau,\mu\tau}\bigr|$ by seeking the minima of their coefficients
in the ${\cal B}(\tau\to e\gamma,\mu\gamma)$ formulas.
Thus for \,$m_{Z'}^{}=150$\,GeV\, the strongest limit we can come up with is
\,$\bigl|b_L^{e\tau}\bigr|<0.2$,\, whereas for \,$m_{Z'}^{}=0.5$~-~2~TeV\,  we get
\begin{eqnarray}
\frac{\bigl|b_L^{e\tau}\bigr|}{m_{Z'}^{}} \,\,\lesssim\,\, 3.3\times10^{-4} ~,
& \hspace{5ex} &
\frac{\bigl|b_R^{e\tau}\bigr|}{m_{Z'}^{}} \,\,\lesssim\,\, 5.5\times10^{-4} ~, \\
\frac{\bigl|b_L^{\mu\tau}\bigr|}{m_{Z'}^{}} \,\,\lesssim\,\, 5.9\times10^{-4} ~,
& \hspace{5ex} &
\frac{\bigl|b_R^{\mu\tau}\bigr|}{m_{Z'}^{}} \,\,\lesssim\,\, 7.4\times10^{-4} ~,
\end{eqnarray}
all of which are less strong than the results in~Eqs.~(\ref{et1}), (\ref{et2}), and~(\ref{mt2})
by about an order of magnitude.
Assuming \,$b_{L,R}^{e\tau}=0$\, and \,$b_{L,R}^{\mu\tau}=0$\, instead leads to, respectively,
\begin{eqnarray}
\frac{\bigl|b_{L,R}^{e\mu}{b}_{R,L}^{\mu\tau}\bigr|}{m_{Z'}^2} \,\,\le\,\, 3.6\times10^{-7} ~,
\hspace{5ex}
\frac{\bigl|b_{L,R}^{\mu e}{b}_{R,L}^{e\tau}\bigr|}{m_{Z'}^2} \,\,\le\,\, 4.2\times10^{-7} ~,
\end{eqnarray}
which are very weak compared to the corresponding constraints deduced from
Eqs.~(\ref{em2}), (\ref{et2}), and~(\ref{mt2})

We remark that these \,$l\to l'\gamma$\, decays effected by the $Z$ and $Z'$, plus additional
loop-induced transitions \,$l\to l'\gamma^*$\, whose amplitudes vanish for a real photon,
also contribute to the flavor-changing decays \,$l\to l'\,\overline{l}{}''l''$.\,
However, due to the loop suppression they are less important than the tree-level
contributions already discussed in Section~\ref{sec:flavorchanging}.

\subsection{Anomalous magnetic moments}

The effective Lagrangian representing the anomalous magnetic moment $a_l^{}$ and electric dipole
moment $d_l^{}$ of a negatively-charged lepton $l$ is
\begin{eqnarray} \label{Lllg}
{\cal L}_{l\bar l\gamma}^{} \,\,=\,\,
\bar l\biggl( \frac{e_p^{}\,a_l^{}}{4m_l^{}}
\,-\, \frac{i\,d_l^{}}{2}\,\gamma_5^{} \biggr) \sigma^{\nu\omega}l\,F_{\nu\omega}^{} ~,
\end{eqnarray}
where \,$F_{\nu\omega}^{}=\partial_\nu^{}A_\omega^{}-\partial_\omega^{}A_\nu^{}$\,
is the photon field-strength tensor.
Nonstandard effects of the $Z$ and $Z'$ on $a_l^{}$ and $d_l^{}$ appear at one-loop level,
arising from the same diagram as in~Fig.\,\ref{loop}, but with \,$l'=l$.\,
From Eqs.~(\ref{Mll'g}) and (\ref{sigmaLR}), we then arrive at the amplitude
\begin{eqnarray} \label{Mllg}
{\cal M}_{l\bar l\gamma}^{} &=&
\frac{i e_p^{}\,\varepsilon_\nu^*k_\omega^{}}{48\pi^2m_Z^2}\sum_j\bar l \Bigl[
3\bigl(\beta_L^{lj}\beta_R^{jl}+\beta_R^{lj}\beta_L^{jl}\bigr) m_j^{} -
2\bigl(\bigl|\beta_L^{lj}\bigr|{}^2+\bigl|\beta_R^{lj}\bigr|{}^2\bigr) m_l^{} +
3\bigl(\beta_L^{lj}\beta_R^{jl}-\beta_L^{jl}\beta_R^{lj}\bigr)m_{j\,}^{}\gamma_5^{} \Bigr]
\sigma^{\nu\omega} l
\nonumber \\ && \! +\; \bigl(Z\to Z',\,\beta\to b\bigr) ~,
\end{eqnarray}
where $k$ is outgoing.
In view of Eq.\,(\ref{Lllg}), the terms without $\gamma_5^{}$ yield
\begin{eqnarray} \label{alZ'}
a_l^{Z'} &=&
\frac{m_l^{}}{12\pi^2\,m_Z^2}\sum_j \Bigl[
3\,{\rm Re}\Bigl(\beta_L^{lj}\beta_R^{jl}\Bigr)m_j^{} -
\Bigl(\bigl|\beta_L^{lj}\bigr|^2+\bigl|\beta_R^{lj}\bigr|^2\Bigr)m_l^{} \Bigr]
\;+\; \bigl(Z\to Z',\,\beta\to b\bigr) ~.
\end{eqnarray}
The same expression can also be derived from Ref.\,\cite{Leveille:1977rc}.
Since the experimental information on $a_\tau^{}$ is still limited~\cite{pdg},
we will address only the \,$l=e$ and $\mu$\, cases.
We then have from Eq.\,(\ref{alZ'})
\begin{eqnarray}
a_e^{Z'} \,=\,\,
\frac{m_e^{}m_\tau^{}\,{\rm Re}\bigl(b_L^{e\tau}b_R^{\tau e}\bigr)}{4\pi^2\,m_{Z'}^2} ~,
\hspace{5ex}
a_\mu^{Z'} \,=\,\,
\frac{m_\mu^{}m_\tau^{}\,{\rm Re}\bigl(b_L^{\mu\tau}b_R^{\tau\mu}\bigr)}{4\pi^2\,m_{Z'}^2} ~,
\end{eqnarray}
where we have kept only the terms proportional to $m_\tau^{}$ and also
neglected terms containing $\beta_L^{l\tau}\beta_R^{\tau l}$.

The SM prediction for $a_e^{}$ agrees with its measurement, their difference being
\,$a_e^{\rm exp}-a_e^{\rm SM}=(-206\pm 770)\times10^{-14}$\,~\cite{Jegerlehner:2009ry}.
On the other hand, the SM and experimental values of $a_\mu^{}$ presently differ by
about 3 sigmas,
\,$a_\mu^{\rm exp}-a_\mu^{\rm SM}=(29\pm9)\times10^{-10}$\,~\cite{Jegerlehner:2009ry}.
Consequently, we may impose
\begin{eqnarray} \label{a_req}
-9.7\times10^{-12} \,\,\le\,\, a_e^{Z'} \,\le\,\, 5.6\times10^{-12} ~, \hspace{5ex}
0 \,\,\le\,\, a_\mu^{Z'} \,\le\,\, 3.8\times10^{-9} ~,
\end{eqnarray}
which translate into
\begin{eqnarray} \label{ammb}
-4.2\times10^{-7} \,\,\le\,\,
\frac{{\rm Re}\bigl(b_L^{e\tau}b_R^{\tau e}\bigr)}
{m^2_{Z'}\;\rm GeV^{{}^{\scriptstyle-2}}}
\,\,\le\,\, 2.4\times10^{-7} ~, \hspace{5ex}
0 \,\,\le\,\,
\frac{{\rm Re}\bigl(b_L^{\mu\tau}b_R^{\tau\mu}\bigr)}
{m^2_{Z'}\;\rm GeV^{{}^{\scriptstyle-2}}}
\,\,\le\,\, 8.0\times10^{-7} ~.
\end{eqnarray}
The result for ${\rm Re}\bigl(b_L^{\mu\tau}b_R^{\tau\mu}\bigr)$ is comparable to that found
in~Ref.\,\cite{Chiang:2006we}.
These bounds are less stringent than those inferred from Eqs.~(\ref{et2}) and~(\ref{mt2}),
respectively.

\subsection{Electric dipole moments}

By comparing the $\gamma_5^{}$ terms in Eqs.~(\ref{Lllg}) and~(\ref{Mllg}), the $Z$ and $Z'$
contributions to the electric dipole moment (EDM) are given by
\begin{eqnarray} \label{dZ'}
d_l^{Z'} \,=\,\, \sum_j^{} \frac{e_p^{}\,m_j^{}}{8\pi^2} \Biggl[
\frac{{\rm Im}\bigl(\beta_L^{lj}\beta_R^{jl}\bigr)}{m_Z^2}+
\frac{{\rm Im}\bigl(b_L^{lj}b_R^{jl}\bigr)}{m_{Z'}^2} \Biggr] ~.
\end{eqnarray}
Obviously, the couplings with \,$j=l$,\, which are real, do not matter in this case.

Since leptonic EDM's have not yet been detected, we will again deal with only
the \,$l=e$ and $\mu$\, cases, the experimental limits on $d_\tau^{}$ being the least restrictive.
We then have from~Eq.\,(\ref{dZ'})
\begin{eqnarray}
d_e^{Z'} \,=\,\,
\frac{e_p^{}\,m_\tau^{}\,{\rm Im}\bigl(b_L^{e\tau}b_R^{\tau e}\bigr)}{8\pi^2\,m_{Z'}^2} ~, \hspace{5ex}
d_\mu^{Z'} \,=\,\,
\frac{e_p^{}\, m_\tau^{}\,{\rm Im}\bigl(b_L^{\mu\tau}b_R^{\tau\mu}\bigr)}{8\pi^2\,m_{Z'}^2} ~,
\end{eqnarray}
where we have neglected terms containing $m_{\mu,e}^{}$ or $\beta_L^{lj}\beta_R^{jl}$.
Since the SM predictions
\begin{eqnarray}
d_e^{\rm SM} \,\,\le\,\, 1\times10^{-38}\,e{\rm\,cm} ~, \hspace{5ex}
d_\mu^{\rm SM} \,\,\le\,\, 3.3\times10^{-25}\,e{\rm\,cm}
\end{eqnarray}
are still negligible compared to the data~\cite{pdg}
\begin{eqnarray}
\bigl|d_e^{}\bigr|_{\rm exp} \,\,\le\,\, 1.6\times10^{-27}\,e{\rm\,cm} ~, \hspace{5ex}
\bigl|d_\mu^{}\bigr|_{\rm exp} \,\,\le\,\, 1.8\times10^{-19}\,e{\rm\,cm} ~,
\end{eqnarray}
we can assume that the latter are saturated by $Z'$ effects.
This translates into
\begin{eqnarray}  \label{edmb}
\frac{\bigl|{\rm Im}\bigl(b_L^{e\tau}b_R^{\tau e}\bigr)\bigr|}{m_{Z'}^2}
\,\,\le\,\, 3.6\times10^{-12} {\rm~GeV}^{-2} ~, \hspace{5ex}
\frac{\bigl|{\rm Im}\bigl(b_L^{\mu\tau}b_R^{\tau\mu}\bigr)\bigr|}{m_{Z'}^2}
\,\,\le\,\, 4.1\times10^{-4} {\rm~GeV}^{-2} ~.
\end{eqnarray}
The first one of these was also evaluated in Ref.\,\cite{Chiang:2006we}, and their result is
roughly similar to ours.
This ${\rm Im}\bigl(b_L^{e\tau}b_R^{\tau e}\bigr)$ constraint appears much stricter than the one
inferred from Eq.\,(\ref{et2}).  But the comparison is actually less clear here due to
the presence of a phase difference between $b_L^{\tau e}$ and $b_R^{\tau e}$ in the former.
In contrast, the ${\rm Im}\bigl(b_L^{\mu\tau}b_R^{\tau\mu}\bigr)$ limit is weaker at least by
3 orders of magnitude than that implied by~Eq.\,(\ref{mt2}).

\section{Predictions \label{sec:predictions}}

We summarize here the strongest limits on the $Z'$ couplings which we have determined.
Defining \,$\hat b_{L,R}^{\ell_i\ell_j}=b_{L,R}^{\ell_i\ell_j}/m_{Z'}^{}$,\,
we have for \,$m_{Z'}^{}=150$\,GeV\,
\begin{eqnarray}  &
-4.7\times10^{-4} \,\,\le\,\, \hat b_L^{ee} \,\,\le\,\,  0.4\times10^{-4} ~, \hspace{5ex}
-6.6\times10^{-4} \,\,\le\,\, \hat b_R^{ee} \,\,\le\,\, -0.6\times10^{-4} ~, & \nonumber \\ &
-2.2\times10^{-4} \,\,\le\,\, \hat b_L^{\mu\mu} \,\,\le\,\, 5.4\times10^{-4} ~, \hspace{5ex}
-2.0\times10^{-4} \,\,\le\,\, \hat b_R^{\mu\mu} \,\,\le\,\, 6.3\times10^{-4} ~, & \nonumber \\ &
-4.6\times10^{-4} \,\,\le\,\, \hat b_L^{\tau\tau} \,\,\le\,\, 1.6\times10^{-4} ~, \hspace{5ex}
     0 \,\,\le\,\, \hat b_R^{\tau\tau} \,\,\le\,\, 5.6\times10^{-4} ~. &
\\ \vphantom{|^{\big|}} &
\bigl|\hat b_L^{e\mu}\bigr| \,\,\le\,\, 3.1\times10^{-7} ~, \hspace{5ex}
\bigl|\hat b_R^{e\mu}\bigr| \,\,\le\,\, 3.8\times10^{-7} ~, & \nonumber \\ &
\bigl|\hat b_L^{e\tau}\bigr| \,\,\le\,\, 1.2\times10^{-4} ~, \hspace{5ex}
\bigl|\hat b_R^{e\tau}\bigr| \,\,\le\,\, 1.5\times10^{-4} ~, & \nonumber \\ &
\bigl|\hat b_{L,R}^{\mu\tau}\bigr| \,\,\le\,\, 1.2\times10^{-4} ~, & \label{ll'1}
\end{eqnarray}
while for \,$m_{Z'}^{}=0.5$~-~2~TeV\,
\begin{eqnarray} & \displaystyle
-5.1\times10^{-4} \,\,\lesssim\,\, \hat b_L^{ee} \,\,\lesssim\,\,
-1.2\times10^{-4} ~, \hspace{5ex}
-5.4\times10^{-4} \,\,\lesssim\,\, \hat b_R^{ee} \,\,\lesssim\,\,
-1.1\times10^{-4} ~, & \nonumber
\\ & \displaystyle
-4.3\times10^{-4} \,\,\lesssim\,\, \hat b_L^{\mu\mu} \,\,\lesssim\,\,
3.4\times10^{-4} ~, \hspace{5ex}
-4.3\times10^{-4} \,\,\lesssim\,\, \hat b_R^{\mu\mu} \,\,\lesssim\,\,
2.1\times10^{-4} ~, & \nonumber
\\ & \displaystyle
-6.1\times10^{-4} \,\,\lesssim\,\, \hat b_L^{\tau\tau} \,\,\lesssim\,\,
-2.0\times10^{-4} ~, \hspace{5ex}
1.9\times10^{-4} \,\,\lesssim\,\, \hat b_R^{\tau\tau} \,\,\lesssim\,\,
5.9\times10^{-4} ~ &
\\ \vphantom{|^{\big|}} &
\bigl|\hat b_L^{e\mu}\bigr| \,\,\lesssim\,\, 1.4\times10^{-7} ~, \hspace{5ex}
\bigl|\hat b_R^{e\mu}\bigr| \,\,\lesssim\,\, 1.8\times10^{-7} ~,
& \nonumber \\ & \displaystyle
\bigl|\hat b_L^{e\tau}\bigr| \,\,\lesssim\,\, 5.3\times10^{-5} ~, \hspace{5ex}
\bigl|\hat b_R^{e\tau}\bigr| \,\,\lesssim\,\, 6.9\times10^{-5} ~,
& \nonumber \\ & \displaystyle
\bigl|\hat b_{L,R}^{\mu\tau}\bigr| \,\,\lesssim\,\, 6\times10^{-5} ~, & \label{ll'2}
\end{eqnarray}
where all the numbers are in units of~GeV$^{-1}$.
The $Z$-pole and LEP-II measurements together have supplied the constraints on
the flavor-conserving couplings.
The numbers for the flavor-changing couplings have come from
\,$\mu\to3e$,\, \,$\tau\to3e$,\, and \,$\tau\to\mu\bar e e$\, data.
In addition, from \,$\mu\to e\gamma$\,
\begin{eqnarray}
\bigl|\hat b_{L,R}^{e\tau}\,\hat b_{R,L}^{\tau\mu}\bigr| \,\,\le\,\, 2.6\times10^{-11}
~\mbox{GeV}^{-2} ~,
\end{eqnarray}
complementary to the individual limits on~$\hat b_{L,R}^{e\tau,\mu\tau}$.
Based on the results above, we now make predictions for the largest values of a number of
observables, including some of those discussed in the preceding two sections.
Our results below can serve the purpose of guiding experimentalists in future searches for $Z'$ signals.

With these couplings, one can obviously get the decay rates of the $Z'$ into a pair of charged
leptons, although not their branching ratios, as we have left its couplings to other fermions
unspecified.
Since \,$\Gamma_{Z'\to\bar l'l}\simeq \bigl(\bigl|b_L^{l'l}\bigr|{}^2+
\bigl|b_R^{l'l}\bigr|{}^2\bigr)m_{Z'}^{}/(24\pi)$,\,
for the flavor-conserving modes we seek values of the couplings which maximize the rates,
but simultaneously satisfy the $Z$-pole and LEP-II requirements discussed
in Section~\ref{sec:flavorconserving}.
For most of the $Z'$ masses considered, the results can roughly be represented by
\begin{eqnarray} \label{Z'll}
\Gamma_{Z'\to e^+e^-}^{} &\lesssim & 7\times10^{-9}~m_{Z'}^3{\rm~GeV}^{-2} ~,
\nonumber \\
\Gamma_{Z'\to\mu^+\mu^-}^{} &\lesssim & 4\times10^{-9}~m_{Z'}^3{\rm~GeV}^{-2} ~,
\nonumber \\
\Gamma_{Z'\to\tau^+\tau^-}^{} &\lesssim & 9\times10^{-9}~m_{Z'}^3{\rm~GeV}^{-2} ~,
\end{eqnarray}
the exceptions being
\,$\Gamma_{Z'\to\mu^+\mu^-}<9\times10^{-9}\,m_{Z'}^3{\rm\,GeV}^{-2}$\, and
\,$\Gamma_{Z'\to\tau^+\tau^-}<6\times10^{-9}\,m_{Z'}^3{\rm\,GeV}^{-2}$\,
in the \,$m_{Z'}^{}=150$\,GeV\, case.
For each of the flavor-violating modes, we simply choose the largest of the relevant set of
$\hat b_{L,R}^{l'l}$ numbers in Eqs.~(\ref{ll'1}) and~(\ref{ll'2}) to arrive at
\begin{eqnarray} \label{Z'll'}
\Gamma_{Z'\to e^\pm\mu^\mp}^{} &\lesssim & 4\times10^{-15}~m_{Z'}^3{\rm~GeV}^{-2} ~,
\nonumber \\
\Gamma_{Z'\to e^\pm\tau^\mp}^{} &\lesssim & 6\times10^{-10}~m_{Z'}^3{\rm~GeV}^{-2} ~,
\nonumber \\
\Gamma_{Z'\to\mu^\pm\tau^\mp}^{} &\lesssim & 4\times10^{-10}~m_{Z'}^3{\rm~GeV}^{-2}
\end{eqnarray}
for \,$m_{Z'}^{}=150$\,GeV\, and their \,$m_{Z'}^{}=0.5$\,-\,2~TeV\, counterparts
with \,$\Gamma_{Z'\to\bar l'l}/m_{Z'}^3$\, ratios which are about 5 times smaller.

Next are the flavor-changing $Z$-boson decays \,$Z\to\bar l'l$.\, Since
\,$\Gamma_{Z\to\bar l'l}\simeq \bigl(\bigl|\beta_L^{l'l}\bigr|{}^2+
\bigl|\beta_R^{l'l}\bigr|{}^2\bigr)m_Z^{}/(24\pi)$\,
and \,$\beta_{L,R}^{l'l}\simeq\xi\,b_{L,R}^{l'l}$,\, we again take for each mode the largest one
of $b_{L,R}^{l'l}$ from Eqs.~(\ref{ll'1}) and~(\ref{ll'2}), but employ the maximal values of $\xi$
consistent with the procedure to determine the couplings in~Section~\ref{sec:flavorchanging}.
Thus, we find that the $b_R^{l'l}$ numbers for \,$m_{Z'}^{}=150$\,GeV\, yield
the largest branching-ratios, namely
\begin{eqnarray} \label{Zll'}
{\cal B}(Z\to e^\pm\mu^\mp) &\le & 4.5\times10^{-12} ~, \nonumber \\
{\cal B}(Z\to e^\pm\tau^\mp) &\le &  6.8\times10^{-7} ~, \nonumber \\
{\cal B}(Z\to\mu^\pm\tau^\mp) &\le &  5.1\times10^{-7} ~.
\end{eqnarray}
The latter two predictions are, respectively, only less than 25 times away from the existing
limits  \,${\cal B}(Z\to e^\pm\tau^\mp)_{\rm exp}^{}<9.8\times10^{-6}$\,  and
\,${\cal B}(Z\to\mu^\pm\tau^\mp)_{\rm exp}^{}<1.2\times10^{-5}$\,~\cite{pdg}.

Turning to the decays of the leptons into 3 lighter leptons, we will address only the modes
that we did not use to derive the strictest constraints.
For \,$\tau\to3\mu$,\, if the upper bounds on $|b_{L,R}^{\mu\tau}|^2$ are used and their
coefficients in the ${\cal B}(\tau\to3\mu)$ expression are maximized, the resulting prediction
for the branching ratio turn out to exceed its experimental limit.
A similar situation arises in~\,$\tau\to e\bar\mu\mu$,\, as can be deduced from its
branching-ratio formula.
Consequently, we cannot make useful predictions in these cases.
Nevertheless, this also means that they may be potential means for probing the~$Z'$
within specific models.
In contrast, for \,$\tau\to e e\bar\mu$\, and \,$\tau\to\bar e\mu\mu$\, we obtain
\begin{eqnarray}
{\cal B}(\tau\to e e\bar\mu) \,\,\le\,\, 1\times10^{-12} ~, \hspace{5ex}
{\cal B}(\tau\to\bar e\mu\mu) \,\,\le\,\, 7\times10^{-13} ~,
\end{eqnarray}
which come from the $\hat b_R^{l'l}$ results for \,$m_{Z'}^{}=150$\,GeV\, and
are much smaller than the current bounds.
The predictions would only double if all the couplings were allowed to contribute
at the same time.
Hence the $Z'$ effects on these 2 modes are unlikely to be detectable in the near future.

The largest impact of the $Z'$ on the effective coupling parametrizing the muonium-antimuonium
conversion is also from the $\hat b_R^{e\mu}$ bound for \,$m_{Z'}^{}=150$\,GeV,
\begin{eqnarray}
|G_C^{}| \,\,=\,\, \frac{\bigl|b_{L,R}^{\mu e}\bigr|^2}{4\sqrt2\;m_{Z'}^2} \,\,\le\,\,
2\times10^{-9}~G_{\rm F}^{} ~,
\end{eqnarray}
far below its experimental counterpart.
Accordingly, we expect that this transition is not sensitive to the $Z'$ signal.

Since the flavor-violating annihilation \,$e^+e^-\to\bar l l'$\, depends on the center-of-mass
energy, we will only give predictions for \,$\bar\sigma(\bar l l')$ at
\,$200{\rm\,GeV}\le\sqrt s\le209$\,GeV\, in the \,$m_{Z'}^{}=150$\,GeV\, case to illustrate
how sensitive these observables might be to the $Z'$ signals.
Thus, searching for the maximal rates, we get
\begin{eqnarray}
\bar\sigma(e\mu) \,\,\le\,\, 6\times10^{-7} {\rm~fb} ~, \hspace{5ex}
\bar\sigma(e\tau) \,\,\le\,\, 0.1 {\rm~fb} ~, \hspace{5ex}
\bar\sigma(\mu\tau) \,\,\le\,\, 0.05 {\rm~fb} ~.
\end{eqnarray}
These numbers are less than the corresponding measured bounds by about 3 orders of magnitude or more.

For the radiative decays, we deal with the rates of \,$\tau\to e\gamma$\, and $\tau\to\mu\gamma$,\,
as \,$\mu\to e\gamma$\, was employed to produce one of the strictest constraints.
Incorporating Eq.~(\ref{Slt}) in (\ref{l2lg}) and dropping the $b_{L,R}^{e\mu}$ terms,
we try to acquire the biggest rates by maximizing the coefficients of $|b_{L,R}^{e\tau,\mu\tau}|{}^2$
in the branching ratios, in a way consistent with the procedure in Section~\ref{sec:flavorchanging}
to extract their upper-limits, and subsequently applying the upper limits, one at a time.
This yields
\begin{eqnarray}
{\cal B}(\tau\to e\gamma) \,\,\le\,\, 2.3\times10^{-8} ~, \hspace{5ex}
{\cal B}(\tau\to\mu\gamma) \,\,\le\,\, 2.1\times10^{-8} ~,
\end{eqnarray}
which are close to the current limits
\,${\cal B}(\tau\to e\gamma)_{\rm exp}^{}<3.3\times10^{-8}$\, and
\,${\cal B}(\tau\to\mu\gamma)_{\rm exp}^{}<4.4\times10^{-8}$\,~\cite{pdg}.

The extent of the $Z'$ contributions to the anomalous magnetic moments and electric dipole
moments of the electron and muon can be learned from~Eqs.~(\ref{ammb}) and~(\ref{edmb}).
Evidently the largest couplings from Eq.\,(\ref{ll'1}) are far from
saturating the maxima of the ranges in~Eq.\,(\ref{ammb}) and the second one
in~Eq.\,(\ref{edmb}), all drawn from comparing the SM expectations and experimental data.
Since the first inequality in~Eq.\,(\ref{edmb}) involves an unknown phase difference between
the couplings, nothing definite can be said of the $Z'$ impact on the electron~EDM in our approach.

Lastly, we would like to make further remarks regarding the situation in the case of no
$Z\mbox{-}Z'$ mixing,~\,$\xi=0$,\, mentioned at the end of Section~\ref{sec:flavorconserving}.
This possibility can arise if one allows the range of the $\rho_0^{}$ parameter from the global
electroweak fit to be slightly enlarged, at 1.14-sigma level to be more precise.
As noted in Section~\ref{sec:flavorconserving}, with \,$\xi=0$\, the flavor-conserving $Z'$
couplings $b_{L,R}^{ll}$ are less constrained than those in the presence of the mixing.
This causes the predictions in Eq.\,(\ref{Z'll}) for  \,$Z'\to\bar ll$\, to rise by about one to
two orders of magnitude.
As another consequence, the steps followed in Section~\ref{sec:flavorchanging} to extract
the strictest limits on the flavor-changing couplings $b_{L,R}^{ll'}$ individually from
\,$\mu\to3e$\, and \,$\tau\to3e,\mu\bar ee$\, are no longer effective, although these decays
could still be useful in restraining products of couplings,
such as $b_{\sf C}^{ee}b_{\sf C'}^{e\mu}$ and~$b_{\sf C}^{\mu e}b_{\sf C'}^{e\tau}$.
The implication is that, with \,$\xi=0$,\, the flavor-changing $Z'$ couplings separately are also
less restricted than in the presence of the mixing, as the bounds now involve only products of
two different couplings, except Eq.\,(\ref{gc}) for~$b_{L,R}^{e\mu}$.
It follows that the predicted number for $\Gamma_{Z'\to e\mu}$ is roughly six orders of
magnitude bigger than that in~Eq.\,(\ref{Z'll'}), whereas the predictions for
$\Gamma_{Z'\to e\tau,\mu\tau}$ can also be expected to be enhanced,
although we cannot be definite about their values.
In the case of \,$Z\to ll'$,\, which proceeds from a~loop diagram if \,$\xi=0$,\,
the enhancement of the branching-ratios in~Eq.\,(\ref{Zll'}) is likely to be modest, if at all,
due to the loop suppression.
For the $\tau$ leptonic and radiative decays, since the products of two different flavor-changing
couplings divided by $m_{Z'}^2$ were calculated in the preceding section to have upper bounds which
are more or less similar, of order $10^{-8}$, the predicted maximum branching-ratios in the absence of
$Z\mbox{-}Z'$ mixing are not far from their experimental limits.

\section{Conclusions \label{sec:summary}}

We have considered a $Z'$ boson with family-nonuniversal couplings to charged leptons and
mixing of kinetic and mass types with the $Z$ boson.
Employing current experimental data and taking a~model-independent approach, we performed
a comprehensive study of constraints on both flavor-conserving and -violating leptonic $Z'$ couplings.
Such an analysis was done for a $Z'$ mass of 150\,GeV, as inspired by recent Tevatron anomalies,
as well as higher masses of \,0.5\,-\,2~TeV.\,
We found that the $Z$-pole and LEP-II measurements together formed the strongest constraints on
the flavor-conserving couplings.  The most stringent bounds on the flavor-changing couplings
came from the measured upper-limits of the branching ratios of the \,$\mu\to 3e$,\,
\,$\tau\to 3e$,\, and \,$\tau\to\mu\bar e e$\, processes.
The radiative decay \,$\mu\to e\gamma$\, supplied complementary information on the flavor-changing
$\mu$-$\tau$ and $e$-$\tau$ couplings.
Detailed results are summarized in the beginning of Section~\ref{sec:predictions}.

With the most restricted of the extracted couplings, we computed the maximum rates of both
flavor-conserving and -changing decays of the~$Z'$ into a pair of charged leptons as functions
of the $Z'$ mass.  We further predicted the rates of flavor-changing \,$Z\to\bar l l'$,
which are not far below the existing measured bounds.
We found that ${\cal B}(\tau\to3\mu)$ or ${\cal B}(\tau\to e\bar\mu\mu)$ are potentially good
observables to probe the $Z'$ within specific models.
In contrast, the rates for \,$\tau\to ee\bar\mu$\, and \,$\tau\to\bar e\mu\mu$\, were calculated
to be too small to be detected in the near future.
Our predictions for ${\cal B}(\tau\to e\gamma)$ and ${\cal B}(\tau\to\mu\gamma)$ are both very
close to their current experimental limits.  We commented that the $Z'$ boson have comparatively
less significant impact on the anomalous magnetic moments and electric dipole moments of
the electron and muon because of the stringent constraints on its couplings.
Finally, we made a number of remarks about how the limits on the couplings and our predictions
might change in the case of no mixing between the~$Z$ and~$Z'$.

Our results could also serve to constrain the rates of other $Z'$-mediated processes involving
both quarks and leptons, such as the \,$B\to X_s l^+l^-$\, and \,$B_s\to l^+l^-$\, decays,
that have been of great interest recently.
This would require extending the analysis to the quark sector.

\medskip

\acknowledgments

This work was supported in part by the National Science Council of R.O.C. under Grants Nos.
NSC-97-2112-M-008-002-MY3, NSC-100-2628-M-008-003-MY4, and NSC-99-2811-M-008-019, and
by the National Central University Plan to Develop First-class Universities and Top-level
Research Centers.

\appendix

\section{Lagrangians with $\bm{Z}$-$\bm{Z'}$ mixing\label{lag}}

The $Z$-$Z'$ mixing scenario considered in this work has been described
in the literature~\cite{Langacker:2008yv,Foot:1991kb}.\footnote{Some specific aspects of
kinetic mixing have been explored in Ref.\,\cite{delAguila:1995rb}.}
We repeat it here using our notation for completeness.

The interaction eigenstates for the neutral fields of the SM gauge group
SU(2)$_L^{}\times$U(1)$_Y^{}$ are, as usual, $W_3^{}$ and $B$, respectively, and
their coupling parameters are $g$ and~$g_Y^{}$.
We denote the gauge boson of the extra Abelian group U(1)$'$ as $C$ and its coupling~$g_C^{}$.
Including kinetic mixing between $B$ and $C$ and mass mixing between $W_3^{}$, $B$, and~$C$,
we obtain the Lagrangian for the kinetic and mass terms after electroweak symmetry breaking as
\begin{eqnarray}
{\cal L}_{\rm km}^{} &=&
-\mbox{$\frac{1}{4}$}W_3^{\nu\omega}W_{3\nu\omega}^{}
- \mbox{$\frac{1}{4}$}B^{\nu\omega}B_{\nu\omega}^{}
- \mbox{$\frac{1}{4}$}C^{\nu\omega}C_{\nu\omega}^{}
- \mbox{$\frac{1}{2}$}\kappa\, B^{\nu\omega}C_{\nu\omega}^{} \,+\,
\mbox{$\frac{1}{2}$}m_W^2\, W_3^2 + \mbox{$\frac{1}{2}$}m_B^2\, B^2 +
\mbox{$\frac{1}{2}$}m_C^2\,C^2
\nonumber \\ && \! -\;
m_W^{}m_B^{}\, W_3^\nu B_\nu^{} - m_W^{}\,{\mu}^{}\,W_3^\nu C_\nu^{} +
m_B^{}\,{\mu}^{}\,B^\nu C_\nu^{}
\nonumber \\ &=&
-\mbox{$\frac{1}{4}$}\,G_{\nu\omega}^{\rm T}\,K\,G^{\nu\omega} \,+\,
\mbox{$\frac{1}{2}$}\,G_\nu^{\rm T}\,M_G^2\,G^\nu ~,
\label{eq:Lkm}
\end{eqnarray}
where the kinetic-mixing parameter obeys \,$|\kappa|<1$\, as required by the positivity of
kinetic energy, the mass-mixing parameter $\mu$ appears when the Higgs field carries a nonzero
U(1)$'$ charge, and
\begin{eqnarray}
m_W^{} \,\,=\,\, \frac{g\,v}{2} ~, \hspace{5ex} m_B^{} \,\,=\,\, \frac{g_Y^{}v}{2} ~,
\hspace{5ex} m_C^2 \,\,=\,\, M_C^2+\mu^2 ~,
\end{eqnarray}
with $v$ being the Higgs vacuum expectation value and the $M_C^{}$ term coming from U(1)$'$
breaking by a SM-singlet scalar field.  Therefore, in the last equality of Eq.~(\ref{eq:Lkm}),
\begin{eqnarray}
G \,= \left(\!\begin{array}{c} B^{\vphantom{|}} \vspace{1ex} \\ W_3^{} \vspace{1ex} \\
C \end{array}\!\right) , ~~~~~
K \,= \left(\begin{array}{ccc} 1 & 0 & \kappa \vspace{1ex} \\ 0 & 1 & 0 \vspace{1ex} \\
\kappa & 0 & 1 \end{array}\right) , ~~~~~
M_G^2 \,= \left(\!\begin{array}{ccc}
m_B^{2^{\vphantom{|}}} & -m_B^{}\,m_W^{} & m_B^{}\,{\mu}^{} \vspace{1ex} \\
-m_B^{}\,m_W^{} & m_W^2 & -m_W^{}\,{\mu}^{} \vspace{1ex} \\
m_B^{}\,{\mu}^{} & -m_W^{}\,{\mu}^{} & m_{C_{\vphantom{o}}}^2 \end{array}\!\right) .
\end{eqnarray}
The kinetic part of ${\cal L}_{\rm km}^{}$ can be put into diagonal and canonical form
via a nonunitary transformation:
\begin{eqnarray}
\tilde T \,\,=\, \left(\begin{array}{ccccc} 1 && 0 && -\kappa/\sqrt{1-\kappa^2}^{\vphantom{|}}
\vspace{1ex} \\ 0 && 1 && 0 \vspace{1ex} \\ 0 && 0 && 1/\sqrt{1-\kappa^2} \end{array}\right) ,
\hspace{5ex}
\tilde T^{\rm T}K\,\tilde T \,\,=\,\, {\rm diag}(1,1,1) ~.
\end{eqnarray}
Employing
\begin{eqnarray} \label{G}
G \,\,=\,\, \tilde T \left(\begin{array}{ccccc} \cos\theta_{\rm W}^{} &&
-\sin\theta_{\rm W}^{} && 0 \vspace{1ex} \\
\sin\theta_{\rm W}^{} && \cos\theta_{\rm W}^{} && 0 \vspace{1ex} \\ 0 && 0 && 1 \end{array}\right) \!
\left(\begin{array}{c} \hat A \vspace{1ex}\\ \hat Z \vspace{1ex} \\
\hat Z' \end{array}\right) , \hspace{5ex}
\sin\theta_{\rm W}^{} \,=\, \frac{m_B^{}}{{M}_Z^{}} ~, \hspace{5ex}
{M}_Z^2 \,\,=\,\, \frac{m_W^2}{\cos^2\theta_{\rm W}^{}} ~,
\end{eqnarray}
with $\theta_{\rm W}^{}$ being the Weinberg angle, leads to
\begin{eqnarray} \label{lkm0}
{\cal L}_{\rm km}^{} \,\,=\,\,
-\mbox{$\frac{1}{4}$}\bigl(\hat A^{\nu\omega}~~~\hat Z^{\nu\omega}~~~\hat Z^{\prime\nu\omega}\bigr)
\left(\begin{array}{c} \hat A_{\nu\omega}^{} \vspace{1ex} \\
\hat Z_{\nu\omega}^{} \vspace{1ex} \\ \hat Z_{\nu\omega}' \end{array}\right)
+\, \mbox{$\frac{1}{2}$}\bigl(\hat A^\nu~~~\hat Z^\nu~~~\hat Z^{\prime\nu}\bigr)
\left(\begin{array}{ccc} 0 & 0 & 0 \vspace{1ex} \\ 0 & {M}_Z^2 & \Delta \vspace{1ex} \\
0 & \Delta  & {M}_{Z'}^2 \end{array}\right)
\left(\begin{array}{c} \hat A_\nu^{} \vspace{1ex} \\ \hat Z_\nu^{} \vspace{1ex} \\
\hat Z_\nu' \end{array}\right) ~,
\end{eqnarray}
where
\begin{eqnarray} \label{mz'}
\Delta \,\,=\,\, \frac{\kappa^{}\,m_B^{}-{\mu}^{}}{\sqrt{1-\kappa^2}}\,{M}_Z^{} ~, \hspace{5ex}
{M}_{Z'}^2 \,\,=\,\,
\frac{m_C^2-2\kappa\,{\mu}^{}\,m_B^{}+\kappa^2\,m_B^2}{1-\kappa^2} ~.
\end{eqnarray}
Hence $\Delta$ contains both kinetic- and mass-mixing contributions, and
\,${M}_{Z'}^{}=m_C^{}$\, in the absence of kinetic mixing, \,$\kappa=0$.\,
Finally, with
\begin{eqnarray}
\left(\begin{array}{c} \hat A \vspace{1ex} \\ \hat Z \vspace{1ex} \\ \hat Z'
\end{array}\right) =
\left(\begin{array}{ccccc} 1 && 0 && 0 \vspace{1ex} \\ 0 && \cos\xi^{} && -\sin\xi^{}
\vspace{1ex} \\ 0 && \sin\xi^{} && \cos\xi^{} \end{array}\right) \!
\left(\begin{array}{c} A \vspace{1ex} \\ Z \vspace{1ex} \\
Z' \end{array}\right) , \hspace{5ex}
\tan(2\xi) \,=\, \frac{2\Delta}{{M}_Z^2-{M}_{Z'}^2} ~,
\end{eqnarray}
one finds in terms of the mass eigenstates
\begin{eqnarray}
{\cal L}_{\rm km}^{} \,\,=\,\,
-\mbox{$\frac{1}{4}$}A^{\nu\omega}A_{\nu\omega}^{}
- \mbox{$\frac{1}{4}$}Z^{\nu\omega}Z_{\nu\omega}^{}
- \mbox{$\frac{1}{4}$}Z^{\prime\nu\omega}Z_{\nu\omega}' \,+\,
\mbox{$\frac{1}{2}$}\,m_Z^2\,Z^2+\mbox{$\frac{1}{2}$}\,m_{Z'}^2\,Z^{\prime2} ~,
\end{eqnarray}
where the eigenmasses $m_Z^{}$ and $m_{Z'}^{}$ are already listed in Eq.\,(\ref{mzmz'}).

The Lagrangian for the interactions of $W_3^{}$, $B$, and $C$ with fermions is
\begin{eqnarray}
{\cal L}_{\rm int}' \,\,=\,\,
-\bigl(g_Y^{}\,J_Y^\lambda~~~g\,J_3^\lambda~~~g_C^{}\,J_C^\lambda\bigr)
\left(\begin{array}{c} B_\lambda^{} \vspace{1ex} \\ W_{3\lambda}^{} \vspace{1ex} \\
C_\lambda^{} \end{array}\right) ~,
\end{eqnarray}
where $J_{Y,3,C}^\nu$ are the currents coupled to the respective fields.
In terms of the fields $\hat A$, $\hat Z$, and $\hat Z'$ defined in~Eq.\,(\ref{G}),
this Lagrangian can be rewritten as
\begin{eqnarray} \label{int}
{\cal L}_{\rm int}' \,\,=\,\,
-e_p^{}\,J_{\rm em}^\lambda\, \hat A_\lambda^{} \,-\, g_Z^{}\, J_Z^\lambda\, \hat Z_\lambda^{} \,-\,
g_{Z'}^{}\, J_{Z'}^\lambda\, \hat Z_\lambda' ~,
\end{eqnarray}
where
\begin{eqnarray} & \displaystyle
e_p^{}\,J_{\rm em}^\lambda \,\,=\,\,
c_{\rm w}^{}\,g_Y^{}\,J_Y^\lambda \,+\, s_{\rm w}^{}\,g\,J_3^\lambda ~, \hspace{5ex}
g_Z^{}\, J_Z^\lambda \,\,=\,\,
c_{\rm w}^{}\,g\,J_3^\lambda \,-\, s_{\rm w}^{}\,g_Y^{}\,J_Y^\lambda ~, &
\nonumber \\  & \displaystyle
g_{Z'}^{}\,J_{Z'}^\lambda \,\,=\,\,
\frac{g_C^{}\,J_C^\lambda}{c_\chi^{}}\,-\,t_\chi^{}\,g_Y^{}\,J_Y^\nu ~, &
\end{eqnarray}
with
\begin{eqnarray}  & \displaystyle
e_p^{} \,\,=\,\, g\,s_{\rm w}^{} \,\,=\,\, g_Y^{}\, c_{\rm w}^{} ~, \hspace{5ex}
g_Z^{} \,\,=\,\, \frac{g}{c_{\rm w}^{}} ~, \hspace{5ex}
c_{\rm w}^{} \,\,=\,\, \cos\theta_{\rm W}^{} ~,  \hspace{5ex}
s_{\rm w}^{} \,\,=\,\, \sin\theta_{\rm W}^{} ~, &
\nonumber \\  & \displaystyle
t_\chi^{} \,\,=\,\, \frac{\sin\chi}{\cos\chi} ~,  \hspace{5ex}
\sin\chi \,\,=\,\, \kappa ~,  \hspace{5ex}
c_\chi^{} \,\,=\,\, \cos\chi \,\,=\,\, \sqrt{1-\kappa^2} ~. &
\end{eqnarray}
In Eq.\,(\ref{L_int}) we reproduce only the part of ${\cal L}_{\rm int}'$
involving $\hat Z$ and~$\hat Z'$.
We note that the field $\hat A_\lambda^{}$ coupled to the electromagnetic current
$J_{\rm em}^\lambda$ is massless, as Eq.\,(\ref{lkm0}) indicates, and hence identical to
the physical photon.

\section{Cross sections of \,$\bm{e^+e^-\to l^+l^-}$\label{csafb}}

From the amplitude in Eq.\,(\ref{Mee2ll}), with each of the propagators now assumed to have
a simple Breit-Wigner form, follows the cross section
\begin{eqnarray}
\sigma(e^+e^-\to l^+l^-) &=& \frac{4\pi\alpha^2}{3\,s} +
\frac{\alpha}{6} \Biggl[
\frac{\bigl(\beta_L^{ee}+\beta_R^{ee}\bigr)\bigl(\beta_L^{ll}+\beta_R^{ll}\bigr)\bigl(s-m_Z^2\bigr)}
{\bigl(s-m_Z^2\bigr){}^2+\Gamma_Z^2m_Z^2} +
\frac{\bigl(b_L^{ee}+b_R^{ee}\bigr)\bigl(b_L^{ll}+b_R^{ll}\bigr)
\bigl(s-m_{Z'}^2\bigr)}{\bigl(s-m_{Z'}^2\bigr){}^2+\Gamma_{Z'}^2m_{Z'}^2} \Biggr]
\nonumber \\ && \! +\;
\frac{\bigl[\bigl(\beta_L^{ee}\bigr){}^2+\bigl(\beta_R^{ee}\bigr){}^2\bigr]
\bigl[\bigl(\beta_L^{ll}\bigr){}^2+\bigl(\beta_R^{ll}\bigr){}^2\bigr]s}
{48\pi\bigl[\bigl(s-m_Z^2\bigr)\mbox{$^2$}+\Gamma_Z^2m_Z^2\bigr]} +
\frac{\bigl[\bigl(b_L^{ee}\bigr){}^2+\bigl(b_R^{ee}\bigr){}^2\bigr]
\bigl[\bigl(b_L^{ll}\bigr)\mbox{$^2$}+\bigl(b_R^{ll}\bigr){}^2\bigr]s}
{48\pi\bigl[\bigl(s-m_{Z'}^2\bigr)\mbox{$^2$}+\Gamma_{Z'}^2m_{Z'}^2\bigr]}
\nonumber \\ && \! +\;
\frac{\bigl(\beta_L^{ee}b_L^{ee}+\beta_R^{ee}b_R^{ee}\bigr)
\bigl(\beta_L^{ll}b_L^{ll}+\beta_R^{ll}b_R^{ll}\bigr)\bigl(s-m_Z^2\bigr)
\bigl(s-m_{Z'}^2\bigr)s}
{24\pi\bigl[\bigl(s-m_Z^2\bigr)\mbox{$^2$}+\Gamma_Z^2m_Z^2\bigr]
\bigl[\bigl(s-m_{Z'}^2\bigr)\mbox{$^2$}+\Gamma_{Z'}^2m_{Z'}^2\bigr]} ~,
\end{eqnarray}
and the forward-backward asymmetry
\begin{eqnarray}
A_{\rm FB} \,\,=\,\, \frac{\sigma_{\rm FB}^{}(e^+e^-\to l^+l^-)}{\sigma(e^+e^-\to l^+l^-)} ~,
\end{eqnarray}
where \,$\alpha=e_p^2/(4\pi)$\, is the fine-structure constant,
$\Gamma_{Z,Z'}^{}$ are the total widths, and
\begin{eqnarray}
\sigma_{\rm FB}^{}(e^+e^-\to l^+l^-) &=&
\frac{\alpha}{8} \Biggl[
\frac{\bigl(\beta_L^{ee}-\beta_R^{ee}\bigr)\bigl(\beta_L^{ll}-\beta_R^{ll}\bigr)\bigl(s-m_Z^2\bigr)}
{\bigl(s-m_Z^2\bigr){}^2+\Gamma_Z^2m_Z^2} +
\frac{\bigl(b_L^{ee}-b_R^{ee}\bigr)\bigl(b_L^{ll}-b_R^{ll}\bigr)
\bigl(s-m_{Z'}^2\bigr)}{\bigl(s-m_{Z'}^2\bigr){}^2+\Gamma_{Z'}^2m_{Z'}^2} \Biggr]
\nonumber \\ && \! +\;
\frac{\bigl[\bigl(\beta_L^{ee}\bigr){}^2-\bigl(\beta_R^{ee}\bigr){}^2\bigr]
\bigl[\bigl(\beta_L^{ll}\bigr){}^2-\bigl(\beta_R^{ll}\bigr){}^2\bigr]s}
{64\pi\bigl[\bigl(s-m_Z^2\bigr)\mbox{$^2$}+\Gamma_Z^2m_Z^2\bigr]} +
\frac{\bigl[\bigl(b_L^{ee}\bigr){}^2-\bigl(b_R^{ee}\bigr){}^2\bigr]
\bigl[\bigl(b_L^{ll}\bigr)\mbox{$^2$}-\bigl(b_R^{ll}\bigr){}^2\bigr]s}
{64\pi\bigl[\bigl(s-m_{Z'}^2\bigr)\mbox{$^2$}+\Gamma_{Z'}^2m_{Z'}^2\bigr]}
\nonumber \\ && \! +\;
\frac{\bigl(\beta_L^{ee}b_L^{ee}-\beta_R^{ee}b_R^{ee}\bigr)
\bigl(\beta_L^{ll}b_L^{ll}-\beta_R^{ll}b_R^{ll}\bigr)\bigl(s-m_Z^2\bigr)
\bigl(s-m_{Z'}^2\bigr)s}
{32\pi\bigl[\bigl(s-m_Z^2\bigr)\mbox{$^2$}+\Gamma_Z^2m_Z^2\bigr]
\bigl[\bigl(s-m_{Z'}^2\bigr)\mbox{$^2$}+\Gamma_{Z'}^2m_{Z'}^2\bigr]} ~,
\end{eqnarray}
the lepton masses having been neglected.
These formulas agree with those in the literature~\cite{Kors:2005uz}.
Here~\,$\beta_{L,R}^{\ell_i\ell_i}=g_{L,R}^{}/c_\xi^{}+t_\xi^{}b_{L,R}^{\ell_i\ell_i}$.\,
In our numerical computation away from \,$s\sim m_{Z,Z'}^2$,\, we set~\,$\Gamma_{Z,Z'}^{}=0$.

\end{document}